\documentclass[preprint,a4paper,showabstract,superscriptaddress,prb]{revtex4-2}

\setcitestyle{super}
\usepackage{multirow}
\usepackage{graphicx}% Include figure files
\usepackage{epstopdf}
\usepackage{dcolumn}% Align table columns on decimal point
\usepackage{bm}% bold math
\usepackage{hyperref}% add hypertext capabilities
\usepackage[utf8]{inputenc} % allow utf-8 input
\usepackage[T1]{fontenc}    % use 8-bit T1 fonts
\usepackage{lmodern}
\usepackage{url}            % simple URL typesetting
\usepackage{booktabs}       % professional-quality tables
\usepackage{amsfonts}       % blackboard math symbols
\usepackage{nicefrac}       % compact symbols for 1/2, etc.
\usepackage{microtype}      % microtypography
\usepackage[version=3]{mhchem}
\usepackage{natbib}
\usepackage{amsmath}
\usepackage{amssymb}
\usepackage{xcolor}
\usepackage[range-units = single,range-phrase = \text{--},separate-uncertainty = true,detect-all]{siunitx}
	\DeclareSIUnit\linepair{lp}
	\DeclareSIUnit\pixels{px}
\usepackage{soul}
\setstcolor{red}
\usepackage{cancel}

\renewcommand\labelenumi{(\roman{enumi})}
\renewcommand\theenumi\labelenumi

\newcommand*{\alex}[1]{\textcolor{black}{#1}}
\newcommand*{\rev}[1]{\textcolor{black}{#1}}

\renewcommand{\vec}[1]{\bm{\mathrm{#1}}}

\newcommand\out{_\mathrm{out}}

\newcommand\rp{\vec{r}_\mathrm{p}}

\begin{document}
%TC:ignore

%\title{Passive Retrieval of the Transmission Matrix\\ for Deep Imaging of a Multiple Scattering Medium}
%\title{Non-invasive Retrieval of the Transmission Matrix  Deep Inside a Multiple Scattering Medium for 3D Imaging with Incoherent Light}
%\title{Deep adaptive focusing inside scattering media with incoherent light}
%\title{Deep passive focusing inside multiple scattering media}
%\title{In-depth transmission matrix in multiple scattering media}
%\title{In-depth transmission of light inside multiple scattering media}
%\title{In-depth focusing of incoherent light inside multiple scattering media}
%\title{In-depth memory effect for incoherent light in multiple scattering media}
\title{Label-free subcellular 3D imaging of oocytes and embryos via reflection matrix microscopy}

\author         {Elsa Giraudat}
\thanks{These authors equally contributed to this work}
\affiliation    {PSL University, ESPCI Paris, CNRS, Institut Langevin, Paris, France}
\affiliation    {OWLO SAS, Paris, France}
\author         {Victor Barolle}
\thanks{These authors equally contributed to this work}
\affiliation    {OWLO SAS, Paris, France}
\author         {Flavien Bureau}
\thanks{These authors equally contributed to this work}
\affiliation    {PSL University, ESPCI Paris, CNRS, Institut Langevin, Paris, France}
\affiliation    {OWLO SAS, Paris, France}
\author         {Nicolas Guigui}
\affiliation    {PSL University, ESPCI Paris, CNRS, Institut Langevin, Paris, France}
\affiliation    {OWLO SAS, Paris, France}
\author         {Paul Balondrade}
\affiliation    {PSL University, ESPCI Paris, CNRS, Institut Langevin, Paris, France}
\affiliation    {OWLO SAS, Paris, France}
\author         {Christine Ho}
\affiliation    {OWLO SAS, Paris, France}
\author         {Vincent Brochard}
\affiliation    {INRAE, BREED, Jouy-en-Josas, France}
\author         {Olivier Dubois}
\affiliation    {INRAE, BREED, Jouy-en-Josas, France}
\author         {Am\'{e}lie Bonnet-Garnier}
\affiliation    {INRAE, BREED, Jouy-en-Josas, France}
\author         {Alexandre Aubry}
\thanks{Corresponding author: alexandre.aubry@espci.fr}
\affiliation    {PSL University, ESPCI Paris, CNRS, Institut Langevin, Paris, France}

\date{\today}
\begin{abstract}
  %  \textbf{High-resolution, label-free imaging is essential for assessing oocyte and embryo \rev{quality} in assisted reproduction. However, dense cytoplasmic lipids induce local refractive index fluctuations that trigger severe optical aberrations and multiple scattering, fundamentally limiting the penetration depth and resolution of conventional microscopy. Here, we introduce an ultra-fast Reflection Matrix Imaging (RMI) platform that captures the electromagnetic field reflected by the sample for a set of \rev{plane wave illuminations at different wavelengths}. By computationally compensating for complex wavefront distortions, RMI realigns forward multiple scattering trajectories with the single-scattering contribution, thereby restoring diffraction-limited 3D resolution (\rev{270} nm) in real-time. We demonstrate non-invasive, sub-cellular visualization of the germinal vesicle through the corona radiata and the inner cell mass within intact blastocysts. By digitally clearing the scattering \rev{mainly}s induced by lipids, RMI overcomes the physical barriers to deep-tissue imaging, providing a robust, label-free tool for objective \rev{oocyte and embryo developmental monitoring in real-time.}}
    \rev{Non-invasive morphological assessment is the cornerstone of oocyte and embryo selection in assisted reproductive technology, yet clinical practice remains limited by two-dimensional, qualitative microscopy. While three-dimensional (3D) fluorescence imaging provides cellular insights, its inherent phototoxicity precludes routine clinical use. Conversely, existing label-free modalities fail to resolve subcellular structures in thick specimens due to two distinct physical barriers: large-scale refractive index heterogeneities, such as the cumulus cells surrounding oocytes, that induce severe aberrations; and short-scale fluctuations, primarily from cytoplasmic lipids, that generate a multiple scattering ``fog''. Here, we report an ultra-fast Reflection Matrix Imaging (RMI) platform designed to overcome these depth and resolution limits. By capturing the back-scattered electromagnetic field for a set of plane-wave illuminations at multiple wavelengths, we record a multi-spectral reflection matrix. From this matrix, we leverage digital adaptive focusing algorithms to computationally compensate for sample-induced aberrations while realigning forward multiple scattering trajectories with the single-scattering contribution. This approach enables label-free 3D visualization of oocytes and blastocysts with an unprecedented subcellular resolution of 300 nm throughout the entire specimen volume. We demonstrate the reliable identification of germinal vesicles and nuclear status in stages previously inaccessible to conventional optics, including imaging through dense cumulus cells. Our method provides a powerful, non-invasive tool for objective grading across all pre-implantation stages, potentially transforming decision-making in clinical IVF.}\end{abstract}
%\keywords{imaging, microscopy, multiple scattering, transmission matrices, reflection matrices}
\maketitle
\newpage

%TC:endignore
%\section{Introduction}
\noindent {\large \textbf{Introduction}} 

Embryo imaging is the cornerstone of modern in-vitro fertilization (IVF), where morphological assessment dictates the selection of embryos with the highest developmental potential~\citep{gardnerBlastocystScoreAffects2000}. Despite its critical role, clinical practice still relies heavily on two-dimensional phase-contrast microscopy~\citep{hoffmanModulationContrastMicroscope1975}, which provides only qualitative and superficial views of complex 3D biological structures. While fluorescence confocal imaging offers high-resolution 3D insights into chromosomal dynamics~\cite{domingo-muelasHumanEmbryoLive2023}, its requirements for exogenous labeling, inherent phototoxicity~\cite{Ottosen2007}, and susceptibility to photobleaching~\cite{Stephens2003} render it unsuitable for routine clinical use. Consequently, a fundamental challenge remains: achieving high-resolution, non-invasive 3D imaging in real-time through thick, highly scattering tissues.

\begin{figure}[ht]
\centering
\includegraphics[width=\textwidth]{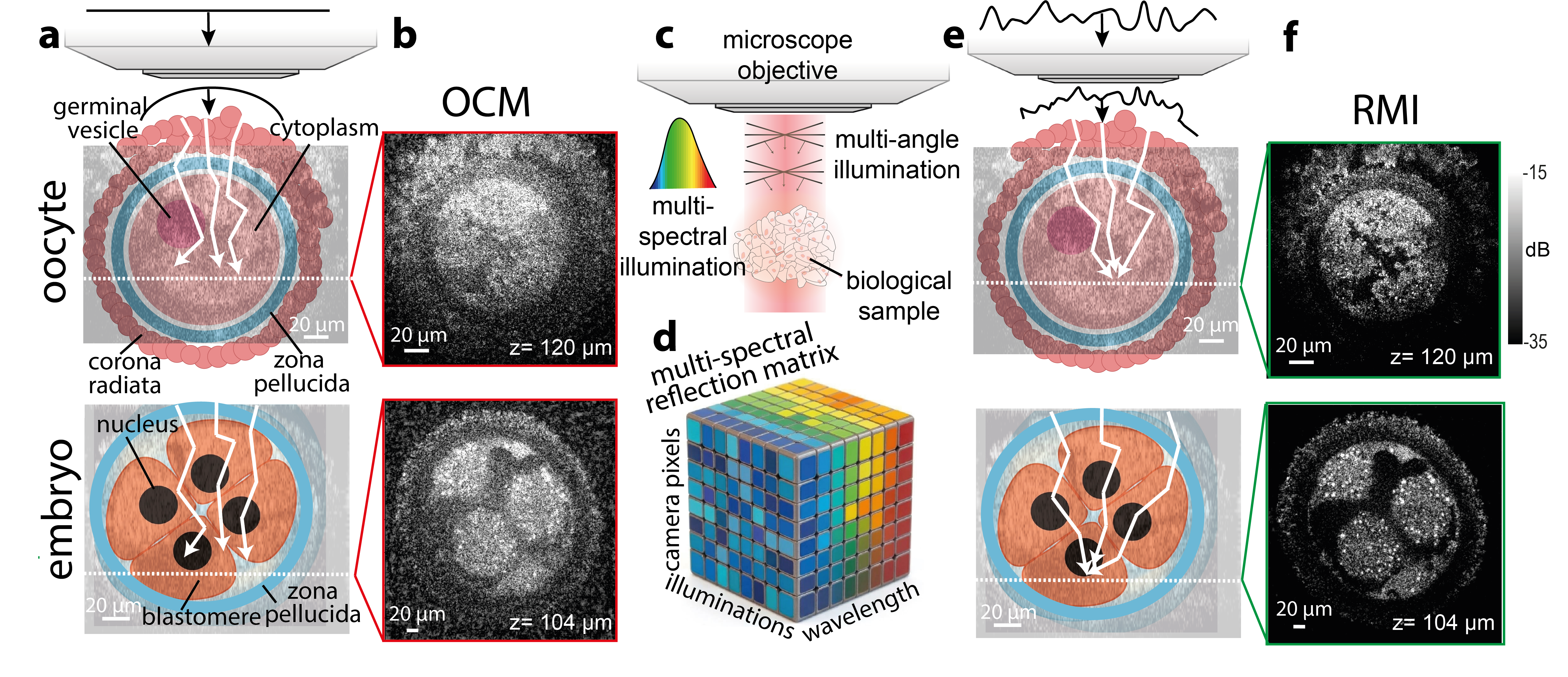}
\caption{\textbf{Fundamental limits of optical coherence microscopy in lipid-rich tissues and the Reflection Matrix Imaging (RMI) solution.} \textbf{a}, Schematic of a cumulus-oocyte complex (top) and four-cell bovine embryo (bottom) overlaid on a longitudinal \rev{$(x,z)$}-section obtained via conventional digital OCM. \textbf{b}, En-face OCM images of the \rev{cumulus-oocyte complex} (top) and embryo (bottom) at depth $z=120$ and 104 $\mu$m, respectively. Forward multiple scattering and aberrations induced by the corona radiata and cytoplasmic lipids (white arrows in a) create a stochastic background that obscures sub-cellular features at depth. \textbf{c}, To overcome this fundamental problem, RMI consists in recording the wave-field scattered by each biological specimen for a discrete set of plane-wave illuminations across a broad spectral range (800-875 nm). \textbf{d}, The resulting multi-spectral reflection matrix contains the complete input-output response of the medium at each \rev{wavelength}. \textbf{e}, Computational post-processing of this matrix rephases distorted wavefront trajectories to restore diffraction-limited resolution. \textbf{f}, This process clears the optical fog observed in (\textbf{b}) and provides high-contrast images of the oocyte (top) and embryo (bottom) internal architecture. Each RMI image has been normalized by its maximum.}
\label{fig0}
\end{figure}

\rev{Current label-free modalities, such as optical diffraction tomography (ODT)~\citep{Haeberle2010,Ling2018,Chowdhury2019} and optical coherence microscopy (OCM)~\citep{Izatt1994,Beaurepaire1998}, face significant physical barriers when applied to clinical specimens. While these techniques can successfully image denudated oocytes~\cite{morawiecFullfieldOpticalCoherence2024}, the cumulus cells surrounding the oocyte generate high-order aberrations (Fig.~\ref{fig0}a) that prevent reliable visualization of germinal vesicle breakdown and polar body extrusion for accurate maturity grading (Fig.~\ref{fig0}b). Evaluation of oocyte maturity is therefore only possible after decoronization (i.e by removing the cumulus cells), which excludes the possibility of in-vitro maturation and consequently discards potential oocytes to be fertilized and ultimately embryos to be transferred. }

\rev{While ODT and OCM can successfully image embryos at early stage of development~\citep{karnowskiOpticalCoherenceMicroscopy2017a,Pierre2022,sobkowiakNumberNucleiCompacted2024}, cells are so compact at the blastocyst stage that an objective evaluation of their number, nuclear status and quality is impossible. Indeed, refractive index variations between cytoplasm, nuclei and extra-cellular matrix induce wave distortions that degrade focusing at large penetration depths (Fig.~\ref{fig0}a). Moreover, cytoplasmic lipid composition induces an optical ``fog'' (Fig.~\ref{fig0}b) that prevents the reliable visualization of deep intracellular structures, such as nuclei and organelles, at depths required for accurate clinical grading. Embryologists are thus looking for high-resolution 3D visualization of internal subcellular structures in a fully non-invasive way, allowing better informed decision-making and ultimately higher efficacy for assisted reproductive technology (ART).  }

Inspired by recent advances in matrix-based wave propagation~\cite{badon_distortion_2020,yoon_laser_2020,Zhang2023,Lee2023}, we report a Reflection Matrix Imaging (RMI) \rev{platform} designed to overcome these fundamental limitations. By integrating a wavelength-swept source (800-875 nm), a sparse illumination sequence (hundreds of plane waves), and ultra-fast holographic detection (1.8 Gb/s), we record the multispectral reflection matrix of the specimen~\cite{balondradeMultispectralReflectionMatrix2024} (Figs.~\ref{fig0}c,d). Leveraging digital adaptive focusing algorithms~\cite{Bureau2023,Najar2023}, a digital clearing of refractive index heterogeneities is performed (Fig.~\ref{fig0}e). Unlike conventional gates that simply reject out-of-focus light, our method computationally realigns the forward multiple scattering trajectories with the single-scattering contribution. This process compensates for sample-induced aberrations and restores an unprecedented subcellular 3D resolution of \rev{300 nm} throughout the entire volume of the oocyte or embryo without any labeling (Fig.~\ref{fig0}f).\\

\noindent {\large \textbf{RESULTS}} 

\noindent {\textbf{Recording the Reflection Matrix}} 

We implemented the measurement of the reflection matrix using a Linnik-type interferometer equipped with dual matched \rev{dry} microscope objectives (numerical aperture,  NA=0.75) in the sample and reference arms~\cite{balondradeMultispectralReflectionMatrix2024}. The imaging procedure (Methods) employs a wavelength-swept laser (tunable from 800 to 875 nm, Extended Data Fig.~\ref{fig1_setup}a) focused onto the objective's pupil plane (Extended Data Fig.~\ref{fig1_setup}b). By scanning this focal spot \rev{in the pupil plane}, we generate a discrete sequence of plane waves that span the sample's angular spectrum (Extended Data Fig.~\ref{fig1_setup}c). This illumination strategy ensures a comprehensive interrogation of the medium's spectral and spatial degrees of freedom \rev{without moving the sample, thereby allowing easy integration in clinics process}. For each incident angle and wavelength, the back-scattered field is captured by a high-speed camera conjugated with the \rev{MO} focal plane, where it interferes with a reference wavefront. The recorded interferograms are organized into a multi-spectral reflection matrix, $\mathbf{R}$, which provides a deterministic representation of the medium's optical response. By leveraging RMI algorithms (Methods), our system computationally compensates for complex wavefront distortions, bypassing the physical limitations of scattering to enable diffraction-limited volumetric imaging of the sample's internal architecture.\\

\noindent {\textbf{Computational Reconstruction and Confocal Imaging}} 

To reconstruct the volumetric data, a numerical focusing procedure is applied to the reflection matrix at each depth $z$. This process involves~\cite{balondradeMultispectralReflectionMatrix2024}: (i) the application of Fresnel propagators to both the input and output dimensions of the matrix to recover axial resolution; \rev{and (ii) a time gating operation to reject multiply-scattered light}. The result is a focused reflection matrix $\mathbf{R}_{\bm{\rho\rho}}(z)=[R(\bm{\rho}_\textrm{out},\bm{\rho}_\textrm{in},{z})]$, which maps the time-gated response between virtual sensors located at transverse positions $\bm{\rho}_\textrm{in}$ and $\bm{\rho}_\textrm{out}$ within the focal plane at depth $z$ (Extended Data Fig.~\ref{ext_fig0}a). An \rev{OCM-like} image $I_C$ is then synthesized \rev{at each depth} by extracting the diagonal elements \rev{of $\mathbf{R}_{\bm{\rho\rho}}(z)$  (\textit{i.e} the time-gated confocal signal)}, where $\bm{\rho}_\textrm{in}=\bm{\rho}_\textrm{out}$:
\begin{equation}
\label{eq5}
I_C(\bm{\rho},z) = R(\bm{\rho},\bm{\rho},z)
\end{equation}
\rev{Unlike full-field OCM that requires a mechanical scan of the sample in depth~\cite{morawiecFullfieldOpticalCoherence2024}, this computational approach exploits the multi-spectral reflection matrix to reconstruct a three-dimensional image in real time.}

\begin{figure}[ht]
\centering
\includegraphics[width=17cm]{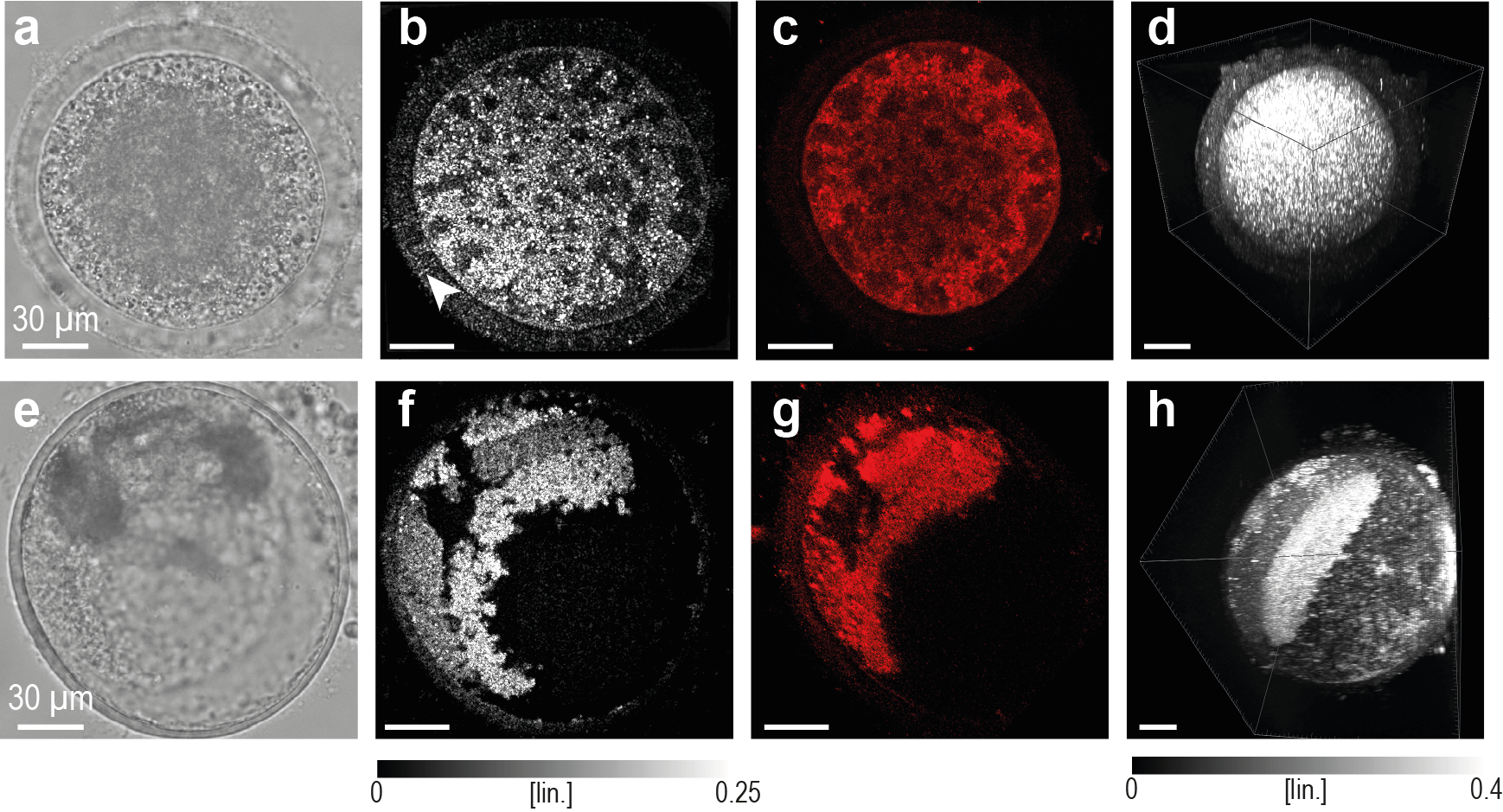}
\caption{\textbf{Origin of speckle in the \rev{oocyte} cytoplasm.} \textbf{a}, Bright-field image of a denuded oocyte. \textbf{b}, Transverse cross-section of the RMI image at depth \rev{$z=70$ $\mu$m highlighting the trans-zonal projections (arrowhead)}. \textbf{c}, Fluorescent lipid image (DiI D-282, see Methods). \textbf{d}, 3D confocal matrix image.  \textbf{e}-\textbf{h}, Same as in panels (a)-(d) after centrifugation of the oocyte. The transverse cross-section of the RMI image in (\textbf{f}) is taken at depth \rev{$z=112$ $\mu$m}. }
\label{fig3}
\end{figure}
\rev{Figure~\ref{fig3} shows the 3D RMI image of a denuded mature oocyte.} The 3D RMI image of a denuded oocyte reveals key morphological features with high fidelity (Fig.~\ref{fig3}b,d). Notably, it resolves the zona pellucida, the trans-zonal projections (TZP, arrowhead in Fig.~\ref{fig3}b) which are critical for intercellular communication~\cite{Clarke2022}, and a highly granular cytoplasm. \rev{This characteristic pattern, often referred to as speckle in wave imaging, a priori result from the interference between randomly distributed unresolved scatterers inside the cytoplasm. Each RMI image has been normalized by its maximum.}

\rev{To unambiguously identify the origin of this speckle pattern,} we performed a density-based stratification via centrifugation (Methods). Histological and epi-fluorescence studies have established that centrifugation segregates \rev{cytoplasmic} oocyte components into five distinct layers: (i) lipids at the centripetal pole, (ii) membrane-bound vesicles, (iii) smooth endoplasmic reticulum, (iv) organelle-free ooplasm, and (v) mitochondria at the centrifugal pole~\cite{Tatham1996,Diez2001}.

\rev{Our 3D reconstruction (Fig.~\ref{fig3}h) and its transverse cross-section (Fig.~\ref{fig3}f)} faithfully recover this stratification. The high-reflectivity granularity aligns perfectly with the lipid-rich pole, a finding corroborated by correlative confocal fluorescence imaging (Figs.~\ref{fig3}c and g). While bright-field microscopy is blinded by the scattering of cytoplasmic lipids (Fig.~\ref{fig3}a and e), RMI computationally clears these distortions. Strikingly, RMI provides superior contrast compared to fluorescence at the center of the oocyte (Figs.~\ref{fig3}b vs. c), where multiple scattering and aberrations typically quench the fluorescent signal. A local compensation of aberrations has actually been performed (Methods, Extended Data Fig.~\ref{ext_fig1}).  This operation is critical to maintain diffraction-limited resolution performance throughout the entire specimen volume. It is even more crucial for imaging an oocyte through cumulus cells as it will be shown now. \\

\noindent {\textbf{Characterization of Spatially-Varying Aberrations and Scattering}} 

\begin{figure}[ht]
\centering
\includegraphics[width=\textwidth]{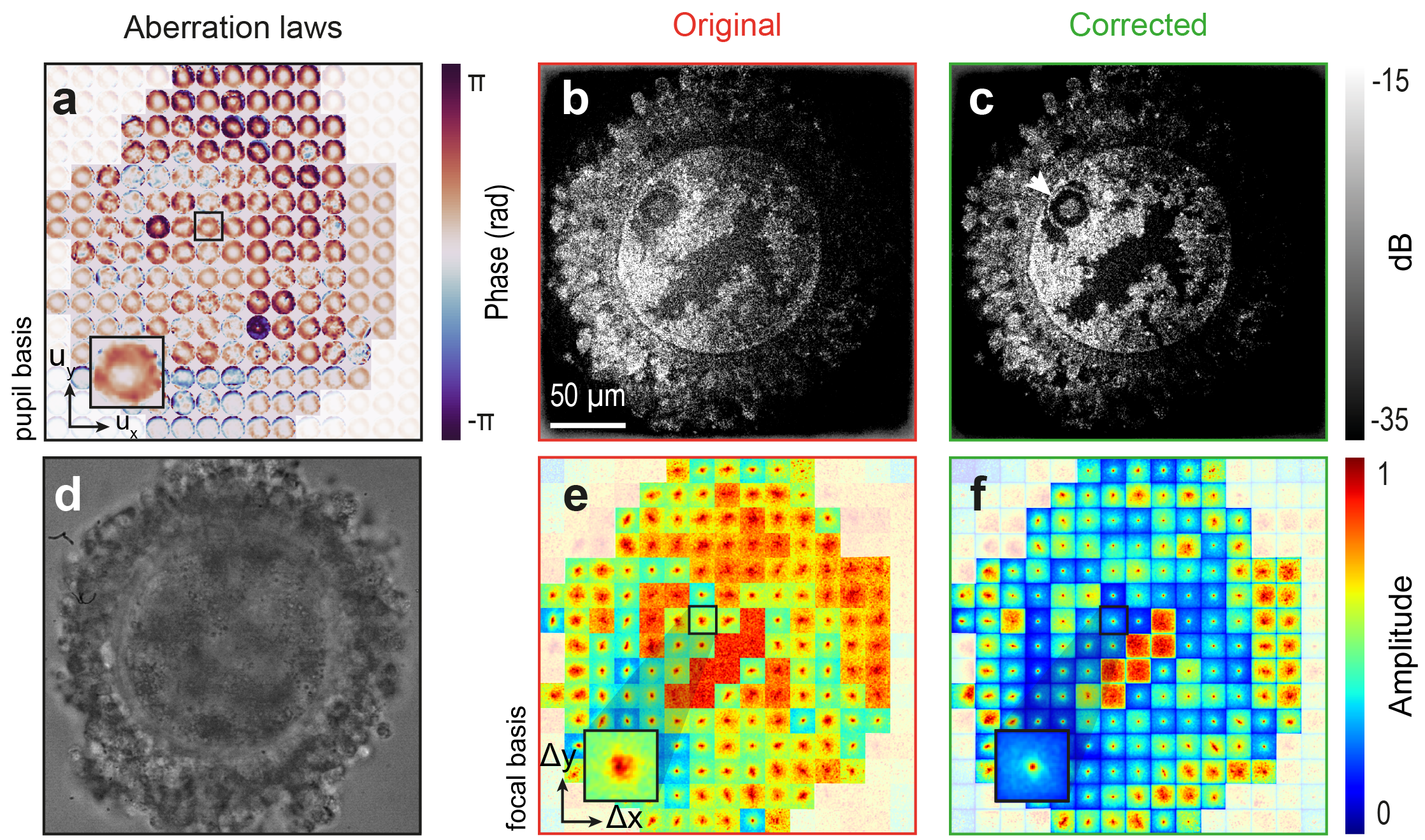}
\caption{\textbf{Reflection matrix imaging of a cumulus-oocyte complex at depth $z=96 $ $\mu$m}. \textbf{a}, Spatial map of the estimated aberration phase laws across the field-of-view. \textbf{b}-\textbf{c}, En-face images of the oocyte (b) before and (c) after the computational aberration correction process \rev{ that unambiguously reveals the germinal vesicle (arrowhead) and its nucleolus}. Each image is normalized by its respective maximum intensity. \textbf{d} Bright-field image of the cumulus-oocyte complex. \textbf{e}-\textbf{f}, Maps of the reflection point-spread function (RPSF) (e) before and (f) after the RMI-based correction, illustrating the restoration of near-diffraction-limited resolution and the suppression of the multiple scattering background.}
\label{fig2_omi}
\end{figure}
\rev{When an oocyte is retrieved, it is surrounded by cumulus cells. In their presence,} traditional bright-field imaging is severely degraded by \rev{aberrations} and multiple scattering (Fig.~\ref{fig2_omi}d). While the axial sectioning and confocal gating inherently provided by RMI significantly enhance image contrast, the reconstruction remains sub-optimal due to persistent aberrations and residual multiple scattering, particularly at greater depths (Extended Data Fig.~\ref{ext_fig2}). RMI uniquely quantifies this degradation by extracting local reflection point-spread functions (RPSFs) from the focused reflection matrix \rev{$\mathbf{R}_{\bm{\rho\rho}}(z)$}. By computationally scanning each virtual detector, $\bm{\rho}_\textbf{out}$, relative to its corresponding virtual source, $\bm{\rho}_\textbf{in}$, the focusing quality can be \rev{locally} characterized across the entire field-of-view (Methods, Eq. \ref{RPSF}).

A representative RPSF map at depth $z=96$ $\mu$m (Fig.~\ref{fig2_omi}e) reveals a characteristic profile: a distorted and broadened confocal peak superimposed upon a diffuse pedestal. While this diffuse background is a hallmark of multiple scattering, the central peak comprises both singly-scattered photons and a coherent backscattering (CBS) contribution arising from constructive interference between reciprocal multiple-scattering paths~\cite{lambert_reflection_2020,Najar2023}.

The spatial fluctuations observed in the RPSF map originate from two primary factors: (\textit{i}) local variations in tissue reflectivity, which modulate the peak-to-background ratio; and (\textit{ii}) transverse refractive index heterogeneities proximal to the focal plane, which induce non-uniform distortions of the confocal spot. This complexity implies that each voxel within the medium is associated with a unique effective focusing law. However, significant spatial correlations persist between RPSFs in adjacent regions (Fig.~\ref{fig2_omi}e), a phenomenon consistent with the principle of optical isoplanatism~\cite{roddier_adaptive_1999} and the generalized memory effect~\cite{osnabrugge_generalized_2017}. This local isoplanicity suggests that aberrations vary slowly over finite sub-regions, providing the physical basis to computationally retrieve local aberration phase laws, $\phi(\mathbf{u},\bm{\rho},z)$, in the pupil plane ($\mathbf{u}$) for every voxel $(\bm{\rho},z)$ throughout the sample volume (Methods, Extended Data Fig.~\ref{ext_fig0}d).\\

\noindent {\textbf{Computational Compensation of Wavefront Distortions}} 

The local aberration phase laws, $\phi(\mathbf{u},\bm{\rho},z)$, are estimated using a multi-scale analysis of wavefront distortions~\cite{Najar2023} (Methods). The resulting aberration map (Fig.~\ref{fig2_omi}a) reveals a complex superposition of spatially-varying, concentric phase patterns and higher-order aberrations that fluctuate across the field-of-view. The concentric patterns correspond to local defocusing induced by refractive index heterogeneities. This low-order aberration is primarily responsible for the broadening of the confocal peak observed in the raw RPSFs (Fig.~\ref{fig2_omi}d). Higher-order aberrations, meanwhile, are linked to forward multiple scattering events that generate the diffuse halos in each RPSF \rev{(see also Extended Data Fig.~\ref{ext_fig2})}.

These estimated phase laws are used to construct the phase of the transmission matrix between the pupil plane and each voxel $(\bm{\rho},z)$ within the specimen volume (Methods). This matrix is then leveraged to deconvolve the focused reflection matrix (Methods), yielding a corrected confocal image (Methods, Fig.~\ref{fig2_omi}c).

Compared to the raw confocal image (Fig.~\ref{fig2_omi}b), the corrected reconstruction exhibits a substantial enhancement in both contrast and spatial resolution. The efficacy of the RMI framework is quantitatively validated by the restored RPSF map (Fig.~\ref{fig2_omi}e). Following computational correction, the confocal peak converges toward the theoretical diffraction limit, $\delta \rho_c \sim \lambda /(4NA) \sim \rev{300}$ nm, across the majority of the field-of-view. The primary exceptions occur within lipid-free regions of the cytoplasm, where the lack of reflectivity results in signal-to-noise ratios insufficient for RPSF characterization.

In the surrounding cumulus oophorus, the RPSF profile reflects a high-order scattering regime. We observe a confocal peak enhancement factor approaching two, a signature we attribute to the coherent backscattering (CBS) effect. This confirms that the confocal signal in this region is predominantly governed by reciprocal multiple scattering paths rather than single-order reflections.

This physical regime accounts for the residual diffuse intensity observed in the confocal image of the cumulus. While RMI successfully compensates for deterministic phase aberrations, the stochastic nature of multiple scattering in the cumulus oophorus maintains a characteristic haze. Conversely, within the oocyte, the diffuse background remains significantly below the confocal peak intensity. In these areas, the signal is dominated by singly-scattered and computationally realigned forward multiply-scattered photons, resulting in a high-contrast reconstruction of the oocyte's structural reflectivity across the objective's full numerical aperture.\\

\noindent {\textbf{Monitoring Oocyte Quality and Maturity through the Corona Radiata}} 

\begin{figure}[ht]
\centering
\includegraphics[scale=1]{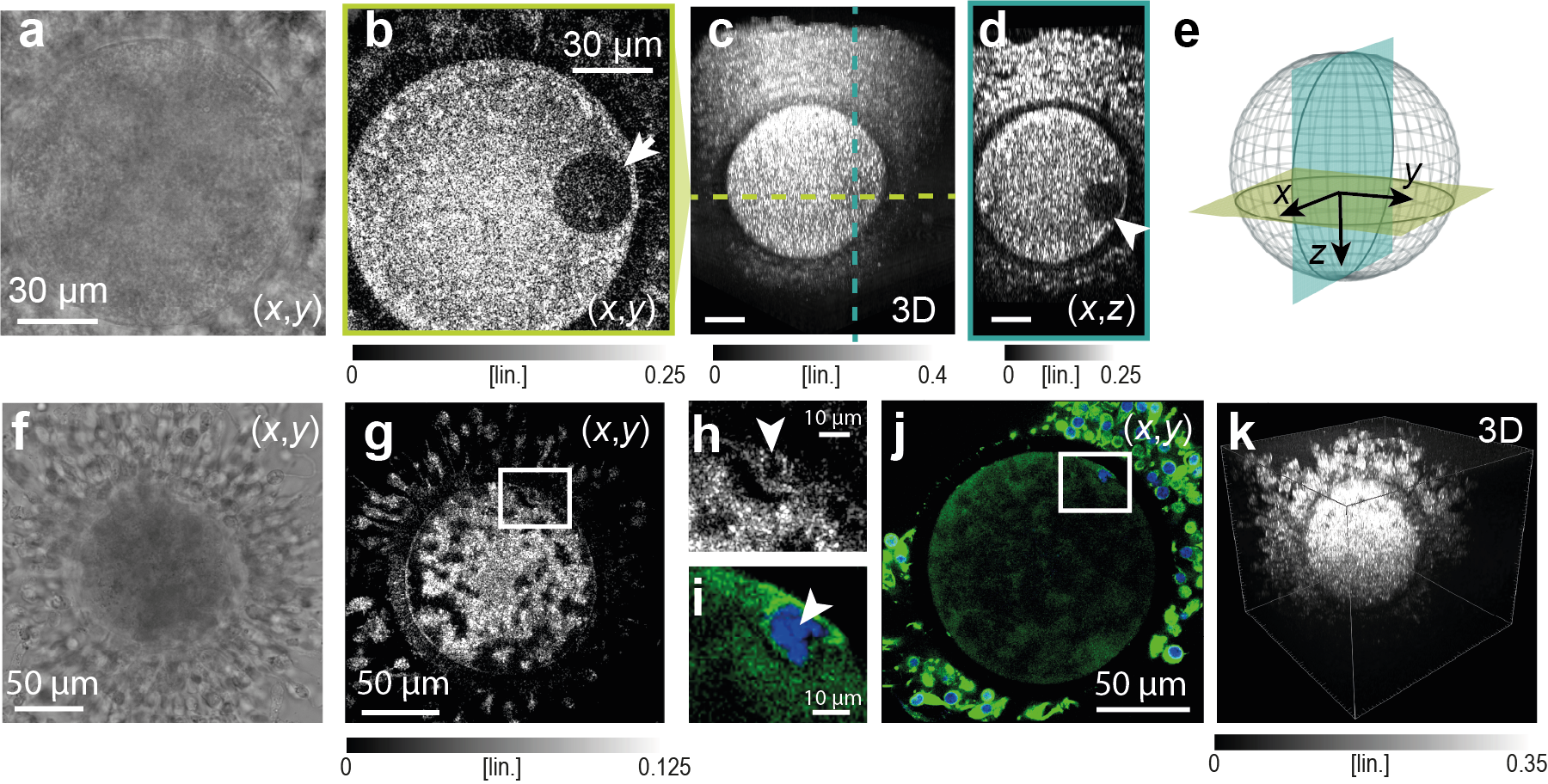}
\caption{\textbf{Three-dimensional imaging of oocytes through cumulus cells.} \rev{\textbf{a}-\textbf{d}, Immature oocyte (scale bar: 30 $\mu$m): \textbf{a}, Bright-field image of the cumulus-oocyte complex; \textbf{b}, Transverse cross-section of the oocyte matrix image at a depth revealing the germinal vesicle (arrowhead) and indicated by a dashed green line in (\textbf{c}); \textbf{c}, 3D view (maximum intensity projection) of the oocyte obtained by RMI; \textbf{d}, Longitudinal cross-section of the oocyte image at a lateral position  revealing the germinal vesicle (arrowhead) and indicated by a dashed cyan line in (\textbf{c}); \textbf{e} 3D representation of the transverse (green) and longitudinal (cyan) cross-sections). \textbf{f}-\textbf{k}, Mature oocyte (scale bar: 50 $\mu$m): \textbf{f}, Bright-field image of the cumulus-oocyte complex. \textbf{g}, Transverse cross-section of the RMI image at a depth revealing the presence of the polar body (white rectangle); \textbf{h}-\textbf{i}, Zoom on the polar body extrusion (arrowhead) corresponding to the white rectangles displayed in the RMI (\textbf{g}) and confocal (\textbf{j}) images (scale bar: 10 $\mu$m); \textbf{j}, Transverse cross-section of the fluorescent confocal image (DNA in blue and alpha-tubulin in green); \textbf{k}, 3D view (maximum intensity projection) of the oocyte obtained by RMI. Each RMI image is normalized by its respective maximum intensity.}}
\label{fig5}
\end{figure}

\rev{After aberration compensation, RMI enables the clear visualization of the germinal vesicle (GV) and its nucleolus within the cytoplasm (Fig.~\ref{fig2_omi}c). It also highlights a bi-phasic distribution of lipids inside the cytoplasm that is abnormal at this development stage of the oocyte. For sake of comparison, the 3D image and the different cross-sections of an healthy immature oocyte are also shown in Figs.~\ref{fig5}b-d. As before, RMI provides a clear visualization of the germinal vesicle (Figs.~\ref{fig5}b,d) unlike the bright-field image (Fig.~\ref{fig5}a). However, in contrast with Fig.~\ref{fig2_omi}, lipids here appear homogeneously distributed throughout the cytoplasm, reflecting a predominant phase of growth and nutrient storage. RMI can therefore evaluate the developmental potential of an oocyte through the cumulus cells by providing the lipid distribution throughout the cytoplasmic volume.}

\rev{Strikingly, RMI can be also a valuable tool to evaluate oocyte maturity through the corona radiata. This capability is demonstrated by the image of a mature oocyte through cumulus cells displayed in Fig.~\ref{fig5}g. Besides highlighting the breakdown of the germinal vesicle, this image shows a distribution of lipids that becomes highly heterogeneous compared to the previous immature case (Fig.~\ref{fig5}c). This observed clustering in mature oocytes aligns with previous reports~\cite{Prates2014,Bradley2016,Khan2021}, suggesting that lipids are reorganized into localized energy reservoirs to be readily mobilized by adjacent mitochondria during the energy-intensive stages of maturation. Interestingly, this image also reveals the presence of the polar body that appears as a small invagination of the oocyte surface inside the white rectangle of Fig.~\ref{fig5}g. A zoom on this area is displayed in Fig.~\ref{fig5}h. To confirm that this modification of the cytoplasmic membrane is the signature of the polar body, immunostaining was used to image DNA (in blue) and alpha-tubulin (in green) on the same oocyte. The corresponding fluorescent confocal image (Fig.~\ref{fig5}j) allow us to identify the polar body colocalized with the invagination (see corresponding inset in Fig.~\ref{fig5}i). Again, none of these characteristic features could have been discerned on the bright-field image in presence of cumulus cells (Fig.~\ref{fig5}f). }

\rev{Label-free matrix imaging can therefore: (\textit{i}) identify the nuclear stage of a coronized oocyte by allowing the detection of germinal vesicle breakdown and polar body extrusion; (\textit{ii}) evaluate its quality by monitoring the lipid distribution throughout its volume. Matrix imaging therefore paves the way towards in-vitro maturation of oocytes, thereby enhancing oocyte fertilization rate and ultimately improving embryo transfer success.} \\

\clearpage

\noindent {\textbf{High-Resolution 3D Mapping of Early-Stage Embryos}} 

\begin{figure}[ht]
\centering
\includegraphics[width=\textwidth]{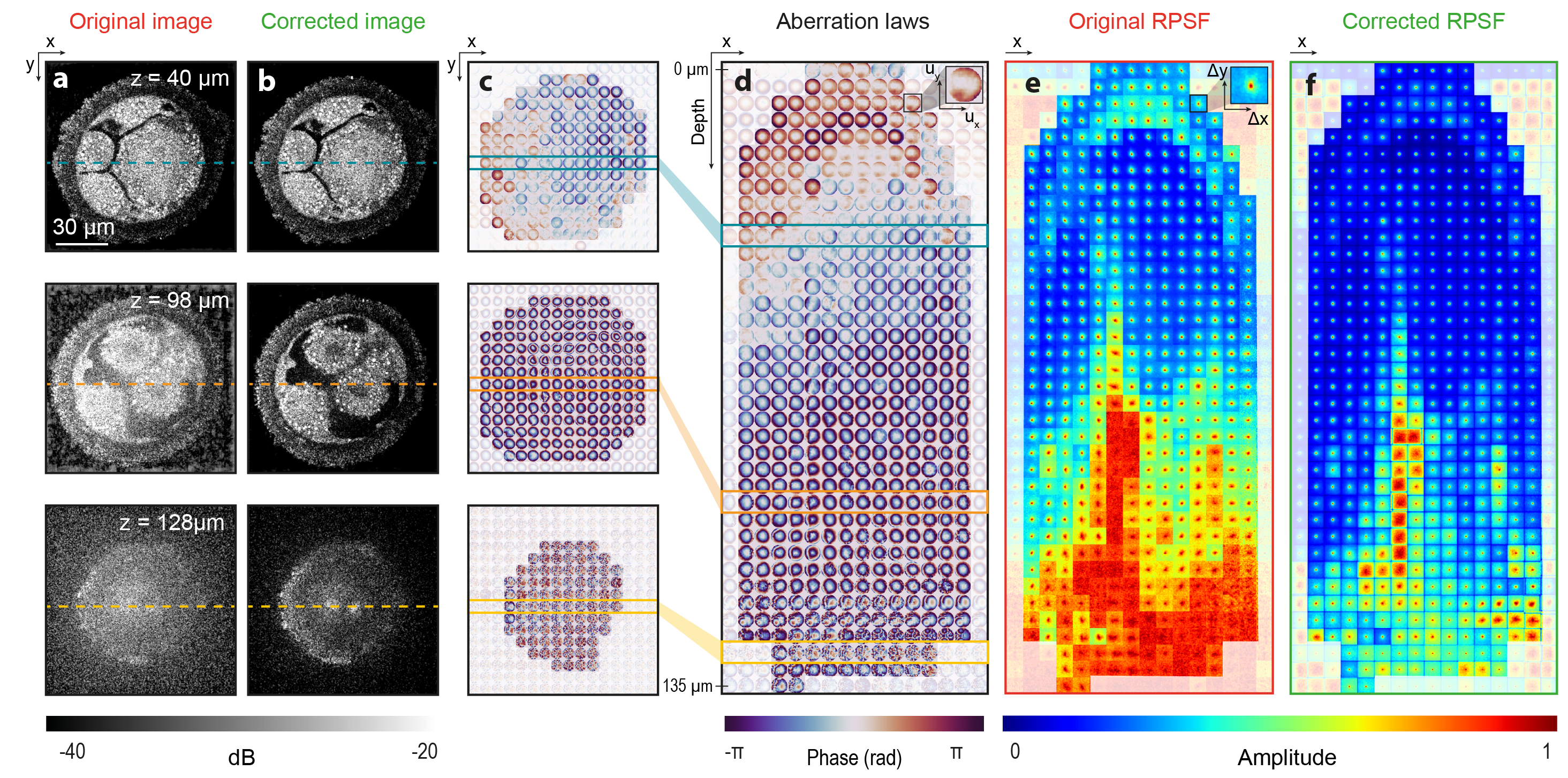}
\caption{\textbf{Reflection matrix imaging of a {Day 2} embryo.} \textbf{a}-\textbf{b}, En-face RMI images before and after aberration compensation at three different depths (scale bar: 30 $\mu$m). Each RMI image has been normalized by its maximum. \textbf{c}, {Map of aberration laws in the corresponding $(x,y)$-plane.} \textbf{d}, {Map of aberration laws in a longitudinal $(x,z)$-plane.} \textbf{e}-\textbf{f}, RPSF map in the same plane, before and after aberration correction, respectively.}
\label{fig6}
\end{figure}

Beyond oocyte \rev{monitoring}, we now demonstrate the capacity of RMI to track the complex morphodynamics of embryonic development at the multicellular stage. As a proof-of-concept, we imaged a bovine embryo at an early developmental stage (Day 2). Similar to our observations in oocytes, bright-field microscopy fails to resolve any internal architecture (Extended Data Fig.~\ref{fig5}). While the confocal and time-gating filters of the RMI framework initially provide sub-cellular insights at shallow depths (Fig.~\ref{fig6}a), image contrast and resolution decrease drastically beyond 70 $\mu$m.

The longitudinal evolution of the RPSF map (Fig.~\ref{fig6}e) confirms this degradation, showing a gradual broadening of the confocal peak and an escalating diffuse background, signatures of an increasing weight of forward multiple scattering trajectories. To realign these paths and restore diffraction-limited quality, we performed a multi-scale analysis of wave distortions. The resulting 3D map of aberration phase laws (Figs.~\ref{fig6}c, d) reveals that while distortions are dominated by simple defocus at shallow depths, their complexity increases significantly with depth. The high spatial-frequency content of $\phi(\mathbf{u})$ in deeper layers is characteristic of the multiple scattering regime encountered in lipid-rich tissues.

Computational compensation via RMI restores a confocal spot whose dimensions approach the diffraction limit while significantly suppressing the multiple scattering background (Fig.~\ref{fig6}f). Although a residual diffuse background persists at extreme depths ($z>120$ $\mu$m, Fig.~\ref{fig6}b bottom panel), likely due to the diffusion limit of the current reconstruction pipeline~\cite{balondradeMultispectralReflectionMatrix2024}, the images are of sufficient fidelity to enable quantitative embryonic analysis.\\

\noindent {\textbf{Label-Free Nuclear and Blastomere Quantification}} \\

Similar to the oocyte, the back-scattering from \rev{lipids and organelles} allows for a clear distinction of individual blastomeres (Fig.~\ref{fig6}a). Strikingly, each cell exhibits a distinct anechoic (dark) inclusion within its cytoplasm. Consistent with previous findings~\cite{karnowskiOpticalCoherenceMicroscopy2017a}, these regions correspond to cell nuclei. This contrast mechanism enables the precise counting of nuclei and cells throughout the embryo volume, a task we also demonstrate in a mouse morula (Extended Data Fig~\ref{fig7}).

The ability to non-invasively assess cell number and identify multi-nucleation events is a critical advancement for clinical IVF. Multi-nucleation is a known marker of chromosomal instability; thus, RMI provides an objective, label-free metric to evaluate an embryo's developmental potential and improve selection criteria~\cite{Coticchio2024}.\\

\noindent {\textbf{Towards Quantitative, Label-Free Morphometry of Embryo}} 

\begin{figure}[ht]
\centering
\includegraphics[width=\textwidth]{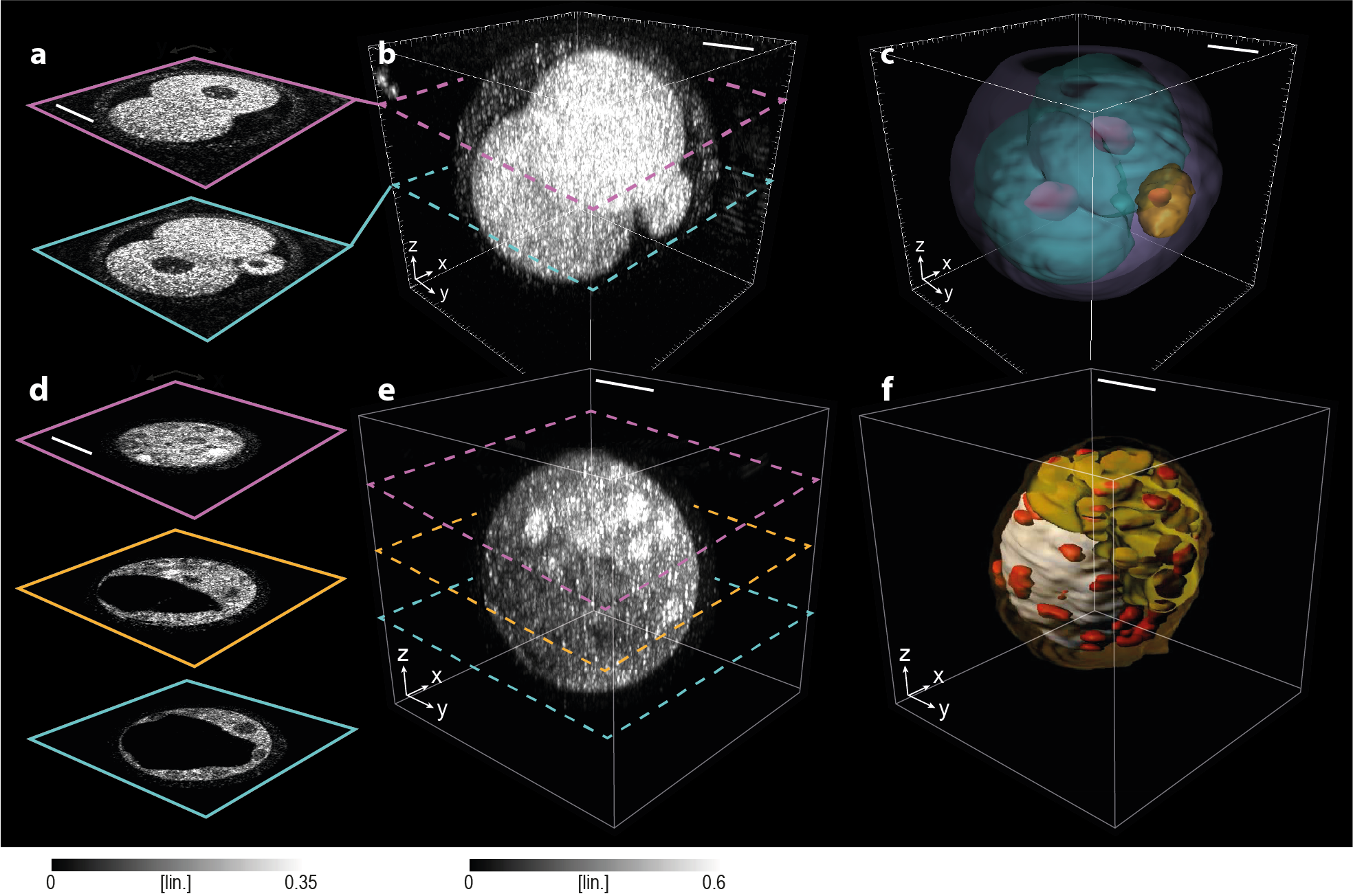}
\caption{\textbf{Embryo segmentation based on reflection matrix images.} \textbf{a}-\textbf{c}, {Day 1} mouse embryo. \textbf{a}, 2D \rev{transverse} cross-sections of the mouse embryo at depth \rev{$z=28$ and 44 $\mu$m}. \textbf{b}, 3D view (maximum intensity projection) of the embryo. \textbf{c}, Segmentation of the different embryo components: Cell cytoplasm in blue, polar body cytoplasm in yellow and nuclei in red. See also Supplementary Movie S1 for a dynamic view of the embryo and its segmentation. \textbf{d}-\textbf{f} Segmentation of a mouse blatocyst.  \textbf{d}, 2D \rev{transverse} cross-section of the mouse blastocyst at depth \rev{$z=22$, 41 and 57 $\mu$m}. \textbf{e}, 3D view (maximum intensity projection) of the blastocyst. \textbf{f}, Segmentation of the different embryo components: Cell nuclei volumes (red), inner cell mass (brown), blastocele (white). See also Supplemntary Movie S2 for a dynamic view of the blastocyst and its segmentation. Each RMI image has been normalized by its maximum.}  
\label{fig5_segmentation}
\end{figure}
As a final proof-of-concept, we demonstrate that RMI enables the precise 3D segmentation of embryonic structures, providing a pathway for objective clinical grading. We first illustrate this on a Day 1 mouse embryo (Fig. \ref{fig5_segmentation}a-c). The high-contrast reflectivity maps allow for the clear delimitation of the two blastomeres (blue, Fig. \ref{fig5_segmentation}c), each containing a single nucleus of significantly lower reflectivity (red). Furthermore, RMI resolves the polar body (brown), a structure often difficult to isolate in 3D without specific markers. This level of detail enables a comprehensive morphological assessment, including the evaluation of cell sphericity, volume, and symmetry. The detection of blastomeres with unequal sizes or irregular volumes can serve as an early indicator of abnormal cleavage, a known marker of reduced embryo quality~\cite{Yao2018}.

The quantitative power of RMI is even more evident at the blastocyst stage (Fig. \ref{fig5_segmentation}d-f). On a Day 5 mouse blastocyst, our segmentation pipeline successfully identifies and counts 32 individual nuclei (red). Moreover, the system differentiates between the dense Inner Cell Mass (ICM, brown), the fluid-filled blastocoel (white), which is characterized by a lack of reflectivity (Fig.\ref{fig5_segmentation}d) \rev{and the trophectoderm where cells are more elongated}.

These results pave the way for a fully automated, label-free quantitative analysis of blastocysts. Critical parameters such as the cell count in the trophectoderm and ICM, the presence of multi-nucleated cells, the volumetric symmetry of compartments, and the cytoplasmic granularity, as well as their morphokinetic evolution, can now be accessed non-invasively. By providing these detailed metrics, RMI offers a standardized, data-driven framework for selecting embryos with the highest implantation potential in IVF clinical workflows.\\

\noindent {\large \textbf{DISCUSSION}} 

Beyond structural segmentation, the RMI framework offers a path toward the precise mapping of physical parameters governing light propagation within the embryo. Parameters such as the local refractive index~\cite{Gul2021} and the scattering mean free path~\cite{Mourant1998} are emerging as crucial, label-free biomarkers for longitudinal developmental monitoring~\cite{MAGATA2023}. As recently demonstrated in the field of ultrasound~\cite{Staehli2023,Heriard2026,Goicoechea2024}, matrix imaging provides a rigorous mathematical foundation~\cite{wasik2026} for extracting these constitutive properties from complex backscattered signals.

Despite the performance of \rev{RMI}, the RPSF measurements (Figs.~\ref{fig2_omi} and~\ref{fig6}) indicate that multiple scattering remains a significant hurdle at extreme depths, particularly for voluminous blastocysts. To further extend the penetration depth, two strategies appear particularly promising. First, the RMI framework is inherently compatible with multi-conjugate adaptive optics strategies, which could more effectively rectify complex forward multiple scattering trajectories by compensating for distortions across multiple axial planes~\cite{Kang2023}. Second, the multi-spectral nature of the reflection matrix provides temporal degrees of freedom that could be exploited to synthesize spatio-temporal focusing laws~\cite{Giraudat2025b}, potentially bypassing the diffusive limit.

The high acquisition speed of \rev{the RMI platform} also paves the way for probing intracellular dynamics. Rapid reflectivity fluctuations (0.1 to 20 Hz) in living specimens are often driven by mitochondrial transport and metabolic activity~\cite{Azzollini2023}. By leveraging the phase stability of RMI, dynamic contrast could enhance the discrimination between nuclei and cytoplasm, identify mitotic states, and detect early signs of cell lysis~\cite{Azzollini2023}. Such metabolic fingerprinting would be invaluable for objective embryo selection. Furthermore, monitoring cell motility offers a label-free alternative to fluorescence, eliminating invasive preparation steps and avoiding the confounding effects of phototoxicity.

In conclusion, this study establishes RMI as a transformative tool for embryology. By delivering high-resolution 3D reconstructions of oocytes and embryos in real-time, RMI surpasses current standards like OCM, which typically rely on slow mechanical scanning and suffer from uncompensated scattering. The \rev{RMI platform} achieves volumetric frame rates an order of magnitude higher than conventional OCM while providing the computational tools necessary to clear the optical distortions inherent to cumulus cells for oocytes and to the complex arrangement of cell and nuclei for blastocysts. We anticipate that RMI will become an essential platform for unraveling the fundamental biological processes of development and significantly improving the predictive accuracy of embryo viability assessments in clinical IVF.\\

%\newpage

\noindent {\large \textbf{METHODS}} \\

\noindent {\textbf{Ethics}} 

Animal care and handling were carried out according to the national rules on ethics and animal welfare in the Animal facility (IERP, INRAE, Infectiology of fishes and rodent facility, doi: 10.15454/1.5572427140471238E12, Jouy-en-Josas, C78-720). This work was approved by the French Ministry of Higher Education, Research and Innovation (n$^{\textrm{o}}$21-01) and the local ethical committee (INRAE Jouy-en-Josas Center). Departmental veterinary regulatory services have delivered habilitation to work with laboratory animals to ABG (n$^{\textrm{o}}$78-184) who supervised the work. \\

\noindent {\textbf{Mouse embryos preparation}} 

Embryos were collected upon superovulation and mating of C57/CBAF1 mice. Adult C57Bl6/CBA F1 female mice were super-ovulated by intraperitoneally injecting 5 IU of pregnant mare serum gonadotropin (PMSG); a second injection, of 5 IU human chorionic gonadotropin (hCG), followed 48 hours later. The mice were sacrificed, and the 1-cell-stage embryos (24 hphCG) were obtained by dissecting the ampulla of the oviduct and treated briefly with 1mg/ml of hyaluronidase in M2 medium (ref Sigma-Aldrich) to remove follicular cells. The harvested embryos were then cultured in vitro in microdroplets of M16 medium (Sigma-Aldrich) under mineral oil (sigma) at 37$^{\circ}$C in a 5\% CO2 atmosphere until they reached the adequate stages.\\

\noindent {\textbf{Bovine oocytes and embryo preparation}} 

\alex{Bovine ovaries were obtained from a local slaughterhouse, washed in NaCl 0.9\% after sampling in NaCl solution 0.9\% and transported in \textit{Euroflush}$^{\textrm{\textregistered}}$  supplemented medium. Cumulus-oocyte complexes (COCs) were collected from each follicle by follicular puncture using a vacuum pump and washed in \textit{Euroflush}$^{\textrm{\textregistered}}$ medium. A stringent selection of COCs was performed to eliminate defective ones before maturation according to the IETS qualitative nomenclature (\textit{International Embryo Technology society}). Subsequently, COCs were cultured in 4-well culture dish per batch of 50. Each pool was put in a well of 500 $\mu$L of medium for 22 hours in a maturation medium (including M199 medium, FCS 10\%, porcine FSH, porcine LH, EGF, estradiol and Gentamycin) under specific humidified atmospheric conditions (5\% CO2 and +38.5$^\textrm{o}$C). After \textit{in vitro} maturation (IVM), MII oocytes were washed with an \textit{in vitro} fertilization (IVF) medium, then put in 4-well culture dish. Simultaneously, frozen semen was thawed and subjected to selection using density \textit{Bovipure}$^{\textrm{\textregistered}}$ gradient centrifugation. Approximately 106 sperm/mL were added to each well of 4-well culture dish containing matured oocytes in IVF medium for 22 hours. Following IVF step at 5\% CO$_2$ at +38.5$^{\circ}$C, cumulus cells were removed by vortexing for 3 minutes, and putative zygotes were sorted and transferred by batch of 30 units to drops of 50 $\mu$L of SOF (\textit{Syntetic Oviductal fluid}) containing essential and non-essential amino acids, sodium pyruvate, BSA and cow's serum at day 3 of estrus. Drops of SOF were covered with paraffin oil and incubated in humidified atmosphere with 5\% CO$_2$ and 5\% O$_2$ at +38.5$^{\circ}$C for further development during 24 to 48 hours. The oocytes (after IVM and removal of the cumulus cells) were either centrifuged 7 min at 13 400g or not before fixation in PFA 2\% and imaged in a drop of PBS. D2 bovine embryo was fixed with 2\% PFA (20 min at RT) prior transfer in a drop of PBS.}\\

%Bovine ovaries were obtained from a local slaughterhouse and transported in NaCl solution 0.9\% . Cumulus-oocyte complexes (COCs) were collected from each follicle and washed in Euroflush® supplemented medium. A stringent selection of COCs was performed to eliminate defective ones before maturation. Subsequently, COCs were cultured for 22 hours in a maturation medium under specific atmospheric conditions (5\% CO2 and 38.5°C). After in vitro maturation (IVM), MII oocytes were washed with an in vitro fertilization (IVF) medium. Simultaneously, frozen semen was thawed and subjected to selection using density bovipure® gradient centrifugation. Approximately 10\textsuperscript{6} sperm/ml were added to each drop containing matured oocytes in IVF medium. Following IVF for 22 hours at 5\% CO2 and 5\% O2 at 38.5°C, cumulus cells were removed by vortexing for 5 minutes, and putative zygotes were transferred to a drop of SOF (Syntetic Oviductal fluid) covered with oil and incubated at 5\% CO2 and 5\% O2 at 38.5°C for further development during 24 to 48 hours.  {D2} and {D3} bovine embryos were fixed with 2\% PFA (20 min at RT) prior transfer in a drop of PBS. The oocytes (after IVM and removal of the cumulus cells) were  either centrifuged 7min at 13 400g or not before fixation in PFA 2\%  and imaging in a drop of PBS.\\
\noindent {\textbf{Oocyte staining}} 

Oocytes were fixed in 2\% paraformaldehyde (PFA) prepared in phosphate-buffered saline (PBS) for 20 min at room temperature (RT). Following fixation, cells were permeabilized in PBS containing 0.5\% Triton X-10 for 20 min at RT. Samples were then incubated for 1 h at 37$^{\circ}$C with the following dyes: BODIPY (1:1000; Invitrogen, D3922, 5 mg/mL stock), MitoBrillant 646 (1:2000; Tocris, Cat. No. 7700, 1 mg/mL stock), and DiI (1:1000; Invitrogen, D282, 2 $\mu$M stock). After staining, oocytes were mounted in small droplets of Vectashield mounting medium on glass-bottom 35 mm dishes (Ibidi). DNA was counterstained with DAPI.  \\

\noindent {\textbf{Oocyte immunolabelling}} 

Oocytes were fixed in PFA2\% (in PBS) for 20min at RT before immunostaining. The oocytes were then permeabilized in 0.5\% Triton X100 / 0.5\% PVP-PBS for 30min at 37$^{\circ}$C and transferred in 2\% BSA-PBS 1h at RT. The oocytes were incubated with primary antibody overnight at 4$^{\circ}$C, washed thrice in 0.5\%PVP-PBS, and then incubated with secondary antibody for 1h. All antibodies were diluted in 2\% BSA-PBS. Antibody against alpha tubulin (1/200; monoclonal Mouse, Sigma T9026) and as secondary antibody an IgG donkey anti-mouse FITC (1/200, 15-095-151, Jackson ImmunoResarch) were used for the immunodetection of tubulin in the bovine oocytes.\\

\noindent {\textbf{Confocal Microscopy}} 

Confocal imaging was performed with a ZEISS LSM 700 confocal laser scanning microscope (MIMA2, INRAE, Microscopy and Imaging Facility for Microbes, Animals and Foods, https://doi.org/10.15454/1.5572348210007727E12) equipped with either a 63X or 40X oil immersion objective. Z-stacks were acquired with a frame size of 512$\times$512 or 1024$\times$1024, a pixel depth of 8 bits, and a z-distance of 0.37 or 0.50 $\mu$m between optical sections. Images were analyzed using Fiji~\cite{Schindelin2012,Schindelin2015,Schneider2012}. \\

\noindent {\textbf{Reflection matrix platform}} 

 The following components were used in the experimental set-up~\cite{balondradeMultispectralReflectionMatrix2024}: A swept laser source (800-875 nm; Superlum-850 HP), 1 galvanometer (Thorlabs, LSKGG4), 2 \rev{dry} objective lenses (Olympus UPLFLN $\times$40; NA 0.75), and a 512$\times$512 ultrafast camera \rev{(5 kHz; Cyclone-1HS-3500)}. \\

\noindent {\textbf{Reflection matrix acquisition}} 

In each RMI experiment, a number $N_{\lambda}$ wavelengths in the $[800-875]$ nm range are scanned with the swept-source laser (Extended Data Tab.~\ref{tab:parameters}). At each wavelength, a number $N_u$ of incident plane waves were used to record the reflection matrix (Extended Data Tab.~\ref{tab:parameters}). Given the magnification of the imaging system and the inter-pixel distance of the camera ($\delta s = {13.7}$ $\mu$m), the output wave field is sampled at a resolution close to $\lambda/(4NA)$, the theoretical achievable resolution for a confocal image: $\delta \rho_c = {300}$ nm.\\

\noindent {\textbf{Data acquisition and graphics processing unit processing}} 
 
All the interferograms of the acquisition sequence are recorded by the camera in 6 s and streamed to the computer's RAM using Bitflow Cyton CXP frame grabbers. The numerical post-processing of the reflection matrix is performed by a graphics processing unit (NVIDIA RTX A6000). For the data sets considered in this paper, all the focusing and aberration correction algorithms are performed in 15 minutes for the entire volume.\\

\noindent {\textbf{Free space focused reflection matrix}} 

The projection of the recorded reflection matrix into the focused basis is performed by applying, in post-processing, focusing operations~\cite{balondradeMultispectralReflectionMatrix2024}. It consists in the application of Fresnel operators by considering water (refractive index $n_0=1.33$) as the propagating medium. The focused reflection matrix is determined over a field-of-view of 15{3}$\times$15{3} $\mu$m$^2$ and a de-scan distance $\Delta \rho={7.5}$ $\mu$m. The spatial sampling is $\delta \rho_c={300}$ nm in the transverse direction and $\delta z=$1 $\mu$m along the $z-$axis. \\

\noindent {\textbf{Reflection point spread functions}} 

To probe the local RPSF, the field-of-view  at each depth $z$ into 16$\times$16 regions defined by their central midpoint $\bm{\rho}_\textrm{p}=(x_\textrm{p},y_\textrm{p})$  and of spatial extension $L=9$ $\mu$m. A local average of the back-scattered intensity is then performed in each region:
\begin{equation}
\label{RPSF}
  RPSF(\Delta \bm{\rho},\bm{\rho}_p,z)=\langle |R(\bm{\rho}_\textrm{out},{\bm{\rho}_\textrm{out}}+\Delta \bm{\rho},z) |^2 W_{L}({\bm{\rho}_\textrm{out}}- \bm{\rho}_p) \rangle_{{\bm{\rho}_\textrm{out}}}
\end{equation}
where the symbol $\langle \cdots \rangle_{m}$ stands for an average over the variable $m$ in subscript. $W_{L}(\bm{\rho}- \bm{\rho}_p) = 1$ for $|x - x_\mathrm{p} |<L/2$ and $|y - y_\mathrm{p} |<L/2$, and zero otherwise.  \\

\noindent \alex{\textbf{Multi-scale compensation of wave-distortions}}

The multi-scale process consists in an iterative compensation of aberration and scattering phenomena at input and output of the reflection matrix. To that aim, wave distortions are analyzed over spatial windows $W_L$ that are gradually reduced at each step $q$ of the procedure, such that: 
\begin{equation}
L=FOV/2^{q-1} 
\end{equation}
where $FOV$ denotes the initial field-of-view.

At each stage of this iterative process, the starting point is the focused reflection matrix $\mathbf{R}^{(q-1)}$, obtained at the previous step, $\mathbf{R}^{(0)}$ being the free space focused reflection matrix. An input distortion matrix $\mathbf{D}_\textrm{in}^{(q)}$ is deduced from $\mathbf{R}^{(q-1)}$ (Extended Data Fig.~\ref{ext_fig0}b,c) via the following matrix operation~\cite{badon_distortion_2020}:
\begin{equation}
\mathbf{D}_\textrm{in}(z)=\left [ \mathbf{R}^{(q-1)}_{\bm{\rho\rho}} (z)  \times \mathbf{P}^T \right ] \circ \mathbf{P}^{\dag}.
\end{equation}
The symbols $\circ$ and $\times$ stand for the Hadamard and matrix product, respectively. The superscripts $^T$ and $^\dag$ stand for matrix transpose and transpose conjugate, respectively. $\mathbf{P}=[P(\mathbf{u},\bm{\rho})]$ is the Fourier transform operator, such that  $P(\mathbf{u},\bm{\rho})=\exp \left [ i2\pi \mathbf{u}\cdot \bm{\rho} /(\lambda f) \right ]$, with $f$ the MO focal length.  

A local correlation matrix of wave distortions is then built around each point ${\mathbf{r}_\textrm{p}=(\bm{\rho}_\textrm{p}},z)$ of the field-of-view (Extended Data Fig.~\ref{ext_fig0}d):
\begin{equation}
\label{corr_in}
    C_\textrm{in} (\mathbf{u}_\textrm{in}, \mathbf{u}'_\textrm{in},\mathbf{r}_\textrm{p})= \left \langle {D}_\textrm{in}({\bm{\rho}\out},z,\mathbf{u}_\textrm{in}){D}_\textrm{in}^{*}({\bm{\rho}\out,z},\mathbf{u}'_\textrm{in}) W_{{L}}(\bm{\rho}\out - \bm{\rho}_\textrm{p})\right \rangle _{{\bm{\rho}\out}}
\end{equation}
Iterative phase reversal~\cite{Bureau2023,Najar2023} is then applied to each correlation matrix $\mathbf{C}_\textrm{in}(\mathbf{r}_\textrm{p})$ (see further). The resulting input phase matrix, $\bm{\Phi}^{(q)}_\textrm{in}(z)=[\phi^{(q)}_\textrm{in}(\mathbf{u},\bm{\rho}_{\textrm{p}},z)]$, is used to compensate for the wave distortions undergone by the incident wave-fronts:
\begin{equation}
    {\mathbf{R}}'_{\bm{\rho\rho}}(z)= \left \lbrace \mathbf{D}_\textrm{in}(z) \circ \exp \left [ -i {\bm{\Phi }}^{(q)T}_\textrm{in} (z) \right] \circ \mathbf{P}^T \right \rbrace \times \mathbf{P}^* 
\end{equation}
where the superscript $^*$ stands for phase conjugate. The corrected matrix $\mathbf{R}'_{\bm{\rho\rho}}$ is only intermediate since phase distortions undergone by the reflected wave-fronts remain to be corrected.

To that aim, an output distortion matrix is built from ${\mathbf{R}}'_{\bm{\rho\rho}}$:
\begin{equation}
 \mathbf{D}_{\textrm{out}}(z)= \mathbf{P}^* \circ \left [ \mathbf{P} \times  \mathbf{R}'_{\bm{\rho\rho}} \right ]
\end{equation}
 From ${\mathbf{D}}_{\textrm{out}}$, one can build a correlation matrix $\mathbf{C}\out$ for each point $\rp$:
\begin{equation}
\label{corr_out}
    C\out (\mathbf{u}\out, \mathbf{u}'\out,\rp)= \left \langle {D}_\textrm{out}(\mathbf{u}\out,\bm{\rho}_\textrm{in} ,{z_p}){D}_\textrm{out}^{*}(\mathbf{u}'\out,\bm{\rho}_{\textrm{in}},{z_p}) W_{L}(\bm{\rho}_{\textrm{in}} - \bm{\rho}_\textrm{p})\right \rangle _{\bm{\rho}_\textrm{in}}
\end{equation}
The iterative phase reversal algorithm described further is then applied to each matrix $\mathbf{C}\out(\rp)$. {The resulting output phase matrix, ${\bm{\Phi}}^{(q)}_\textrm{out}(z)=[\phi^{(q)}_\textrm{out}(\mathbf{u},\bm{\rho}_{\textrm{p}},z)]$, are leveraged to compensate for the wave distortions undergone by the reflected wave-fronts:
\begin{equation}
    \mathbf{R}_{\bm{\rho \rho}}^{(q)}(z)=\mathbf{P}^{\dag} \times  \left \lbrace \mathbf{P} \circ \exp \left [ -i \bm{\Phi}^{(q)}_\textrm{out} (z) \right] \circ \mathbf{D}_\textrm{out} (z) \right \rbrace
\end{equation}
This matrix $\mathbf{R}^{(q)}$ is the starting point of the next stage of the multi-scale process, and so on.}

The final image of each sample (Figs.~\ref{fig2_omi}c and \ref{fig6}b) can be obtained by considering the diagonal elements of the corrected matrix $\mathbf{R}^{(Q)}_{\bm{\rho \rho}}$ at the final step $Q$ of the aberration correction process:
\begin{equation}
\label{eq9}
I_M(\bm{\rho},z)=R^{(Q)}(\bm{\rho},\bm{\rho},z).
\end{equation} 
For each matrix image displayed in the paper, the number of iterations used for the aberration correction process is $Q=5$. The corrected RPSFs displayed in Figs.~\ref{fig2_omi}f and \ref{fig6}f  are extracted from the final matrix $\mathbf{R}^{(Q)}$ using Eq.~\ref{RPSF}. The  aberration phase laws displayed in Figs.~\ref{fig2_omi}a and \ref{fig6}c,d cumulate the output aberration phase laws determined at each step:
\begin{equation}
\bm{\Phi}_{\textrm{in/out}}(z)= \sum_{q=1}^{Q}  \bm{\Phi}^{(p)}_{\textrm{in/out}}(z)
\end{equation}
\\

\noindent {\textbf{Iterative phase reversal algorithm.}} 

The iterative phase reversal algorithm~\cite{Bureau2023} is a computational process that provides an estimator of the {transmittance matrix}, that links each point $\mathbf{u}$ of the pupil plane with each voxel $\rp$ of the sample. To that aim, the correlation matrix $\mathbf{C}$ computed over the spatial window $W_{L}$ centered around each point $\rp$ is considered {(Eqs.~\ref{corr_in} and ~\ref{corr_out})}.  Mathematically, the algorithm is based on the following recursive relation:
\begin{equation}
{{\bm{\Phi}}_n (\rp) = \mathrm{arg}\left\{  \mathbf{C}(\rp) \times \exp \left [ i {\bm{\Phi}}_{n-1}(\rp) \right ]  \right\}}
\end{equation}
where ${\bm{\Phi}}_n$ is the estimator of {${\bm{\Phi}}$} at the $n^\textrm{th}$ iteration of the phase reversal process. \alex{$ {\Phi}_0$} is an arbitrary wave-front that initiates the process (typically a flat phase law) and ${{\bm{\Phi}}}= \lim_{n\to\infty} \alex{{\bm{\Phi}}}_{n}$ is the result of the iterative phase reversal process. \\

%\noindent {\textbf{Data availability}} \\
%\textcolor{red}{Optical data used in this manuscript have been deposited at \alex{Zenodo}.}
%\\

%\noindent {\textbf{Code availability}}
%\\
%Codes used to post-process the optical data within this paper are available upon request at Zenodo~\cite{Balondrade_2024_code} (\href{https://zenodo.org/records/10674114}{https://zenodo.org/records/10674114})\\

\clearpage
\bibliography{embryo_paper}

\newpage

\noindent {\large{\textbf{Acknowledgments}}} \\

EG, VBa, FB, NG, PB, and AA are grateful for the funding provided by the European Research Council (ERC) under the European Union's Horizon 2020 research and innovation program (grant agreements no. 819261 and 101158062, REMINISCENCE and MUSE projects). This project has also received funding from Labex WIFI (Laboratory of Excellence within the French Program Investments for the Future; ANR-10-LABX-24 and ANR-10-IDEX-0001-02 PSL*), {from CNRS Innovation (Prematuration program, MATRISCOPE project)} and from PSL (Prematuration program, MIRE project). VBr, OD, AJ and ABG acknowledge the embryo production facility of the BREED Unit. Their work is funded by the REVIVE Labex (Investissement d'Avenir, ANR-10-LABX-73) and supported by the PHASE Department of the French National Research Institute for Agriculture, Food and Environment (INRAE).\\

\noindent {\large{\textbf{Author contribiutions}}} \\

\alex{PB, VBa, NG, ABG and AA initiated the project. VBr, OD, CH and ABG prepared the samples. VBr, OD, CH and ABG acquired the bright-field and fluorescence images. PB, VBa, and FB built the RMI platform. EG and FB performed the RMI experiments. NG and VBa developed the acquisition software. VBa, EG and FB developed the post-processing tools. EG, NG, PB, ABG and AA analyzed the experimental results. EG, FB and AA prepared the figures. AA, EG, and ABG prepared the manuscript. EG, VBa, FB, NG, PB, CH, ABG and AA discussed the results and contributed to finalizing the manuscript.}\\

\noindent {\large{\textbf{Competing interests}}} \\

\alex{VBa, PB and AA are named inventors on french patent FR2207334 (filing date 18.07.2022), which is related to the techniques described in this Article. PB, VBa, NG and AA are co-founders of the OWLO company, that will commercialize the matriscope used in this study. EG and FB are currently employees of the OWLO company. All authors declare that they have no other competing interests.}\\
\clearpage

%\noindent \textbf{Figure captions.}\\ 

%\bibliographystyle{apsrev4-2}

\noindent \textbf{Extended Data Table.}\\

\setcounter{figure}{0}
\renewcommand{\figurename}{Extended Data Tab.}

\begin{figure*}[ht]
\centering
	\begin{tabular}{|l|l| c|c|} % four columns, alignment for each
	\hline
		Sample & Figures &  $N_{\lambda}$  &  $N_{u}$  \\
		\hline
		Biphasic bovine COC & Fig.~\ref{fig0} (top), Fig.~\ref{fig2_omi}, Extended Data Fig.~\ref{ext_fig2}  & $180$ & $325$ \\
				\hline
		Denuded bovine oocyte &  Fig.~\ref{fig3}, Extended Data Fig.~\ref{ext_fig1} & $120$ & $325$ \\
				\hline
		Immature bovine COC & Fig.~\ref{fig5} (top) & $120$ & $465$ \\
				\hline
		Mature bovine COC & Fig.~\ref{fig5} (bottom) & $120$ & $325$ \\
				\hline
		Four-cell bovine embryo & Fig.~\ref{fig0} (bottom), Fig.~\ref{fig6},  Extended Data Fig.~\ref{fig5ext}  & $120$ & $325$ \\
				\hline
		Two-cell mouse embryo & Fig.~\ref{fig5_segmentation} (top) & $120$ & $149$ \\
				\hline
	    Mouse blastocyst & Fig.~\ref{fig5_segmentation} (bottom) & $120$ & $465$ \\
	    		\hline
	   Mouse morula & Extended Data Fig.~\ref{fig7} & $120$ & $325$  \\
		\hline
	\end{tabular}
\caption{\label{tab:parameters} \textbf{Acquisition parameters.}
		Number of wavelengths ($N_{\lambda}$) and incident plane waves ($N_{u}$) used for each reflection matrix acquisition. }
\end{figure*}

\clearpage

\noindent \textbf{Extended Data Figures.}\\ 

\setcounter{figure}{0}
\renewcommand{\figurename}{Extended Data Fig.}

\begin{figure}[ht]
\centering
\includegraphics[scale=0.7]{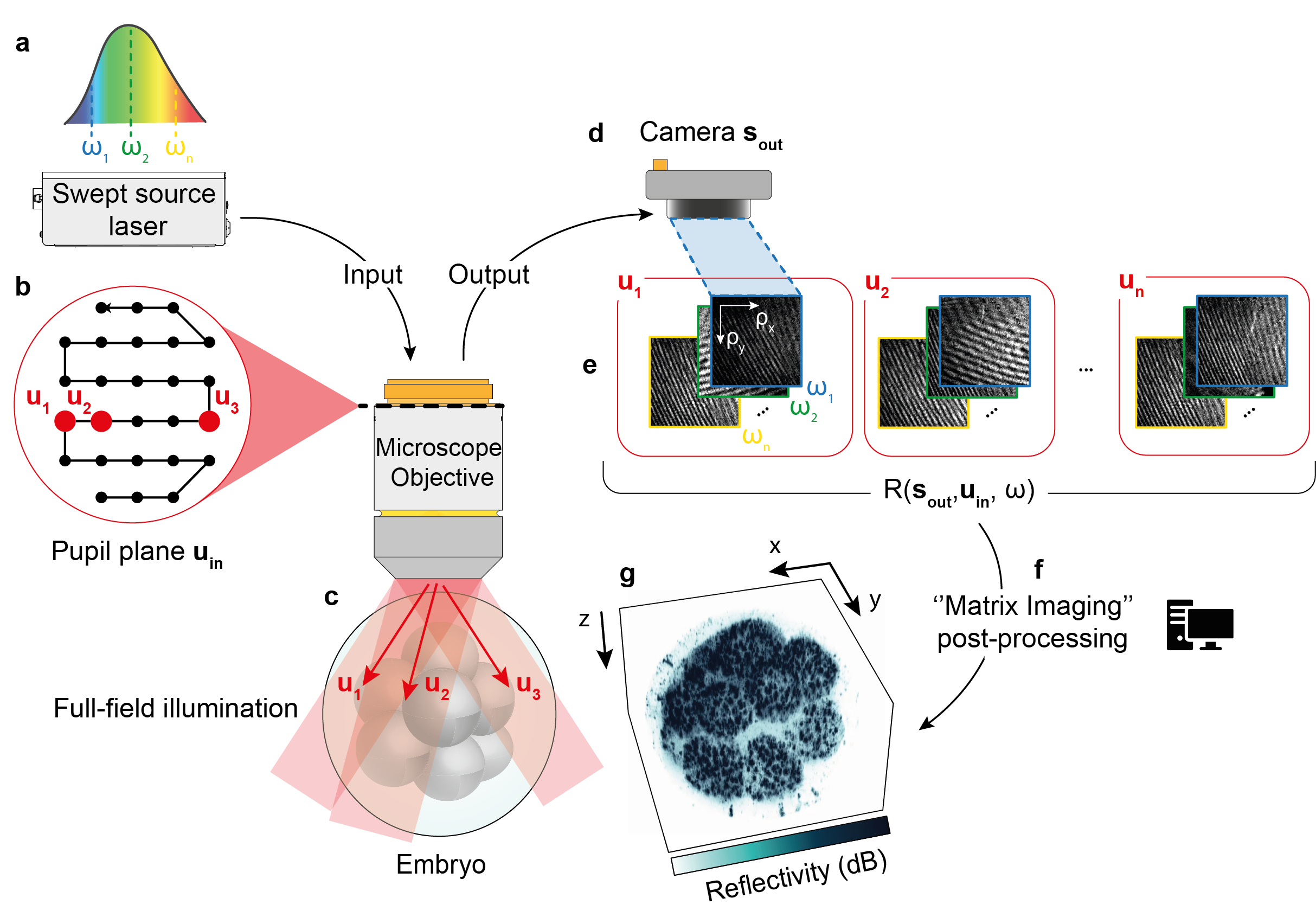}
\caption{\textbf{Reflection Matrix Imaging (RMI) workflow.} \textbf{a}-\textbf{b}, A wavelength-swept laser source (\textbf{a}) is focused at successive positions $\mathbf{u}_\text{in}$ in the microscope objective (MO) pupil plane (\textbf{b}). \textbf{c},Each focal spot in the pupil plane generates a plane wave illuminating the sample at a specific incident angle. \textbf{d}, The back-scattered field interferes with a reference wavefront (not shown) on a high-speed camera conjugated with the sample's focal plane. \textbf{e}, The resulting interferograms are stacked into a multi-spectral reflection matrix, $\mathbf{R}$, capturing the sample's response across angular and spectral degrees of freedom.  \textbf{f},  Post-processing via RMI algorithms (Methods, Extended Data Fig.~\ref{ext_fig0}) compensates for both system and sample-induced aberrations. \textbf{g}, It yields a  3D reconstruction of the specimen at a sub-micron resolution.}
\label{fig1_setup}
\end{figure}

\begin{figure*}[ht]
\centering
\includegraphics[width=\textwidth]{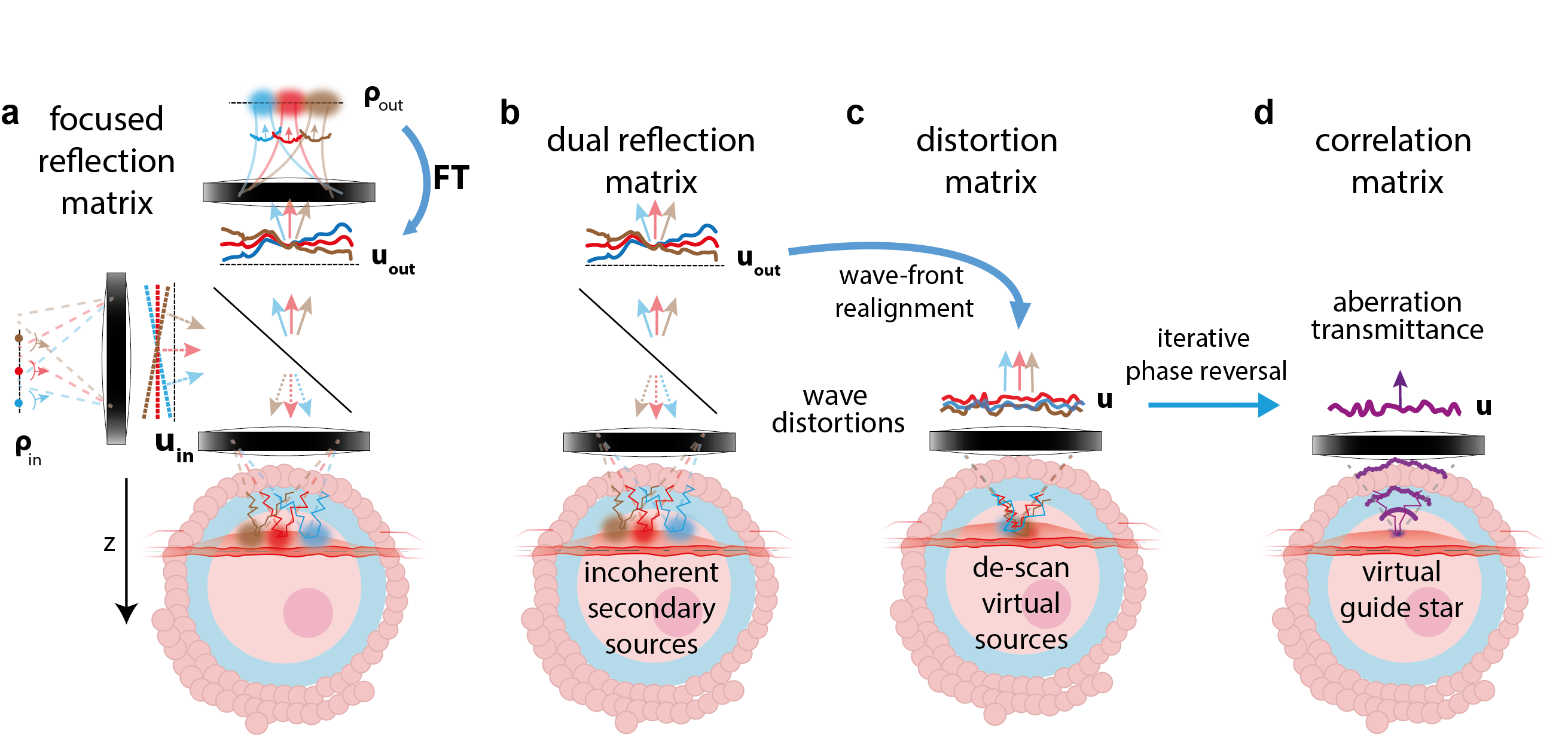}
\caption{\label{ext_fig0}\textbf{Matrix compensation of transverse aberrations.} \textbf{a}, The digitally
refocused matrix $\mathbf{R}_{\bm{\rho \rho}}$ contains the set of impulse responses $R(\bm{\rho}_{\textrm{out}},\bm{\rho}_{\textrm{in}}, z)$ between
an array of point sources $\bm{\rho}_{\textrm{in}}$ and detectors $\bm{\rho}_{\textrm{in}}$ lying in the coherence plane at depth $z$. \textbf{b}, Reflected wave-fronts are projected back in the pupil plane. \textbf{c}, Wave distortions are then isolated by realigning the reflected wave-fronts. Seen from the focal plane, this operation amounts to a de-scan of each focal spot. \textbf{d}, An iterative phase reversal process applied to those wave distortions provides a estimation of local aberration transmittances by recombining each focal spot into a virtual guide star over reduced spatial windows.}
\end{figure*} 

\begin{figure*}[ht]
\centering
\includegraphics[width=\textwidth]{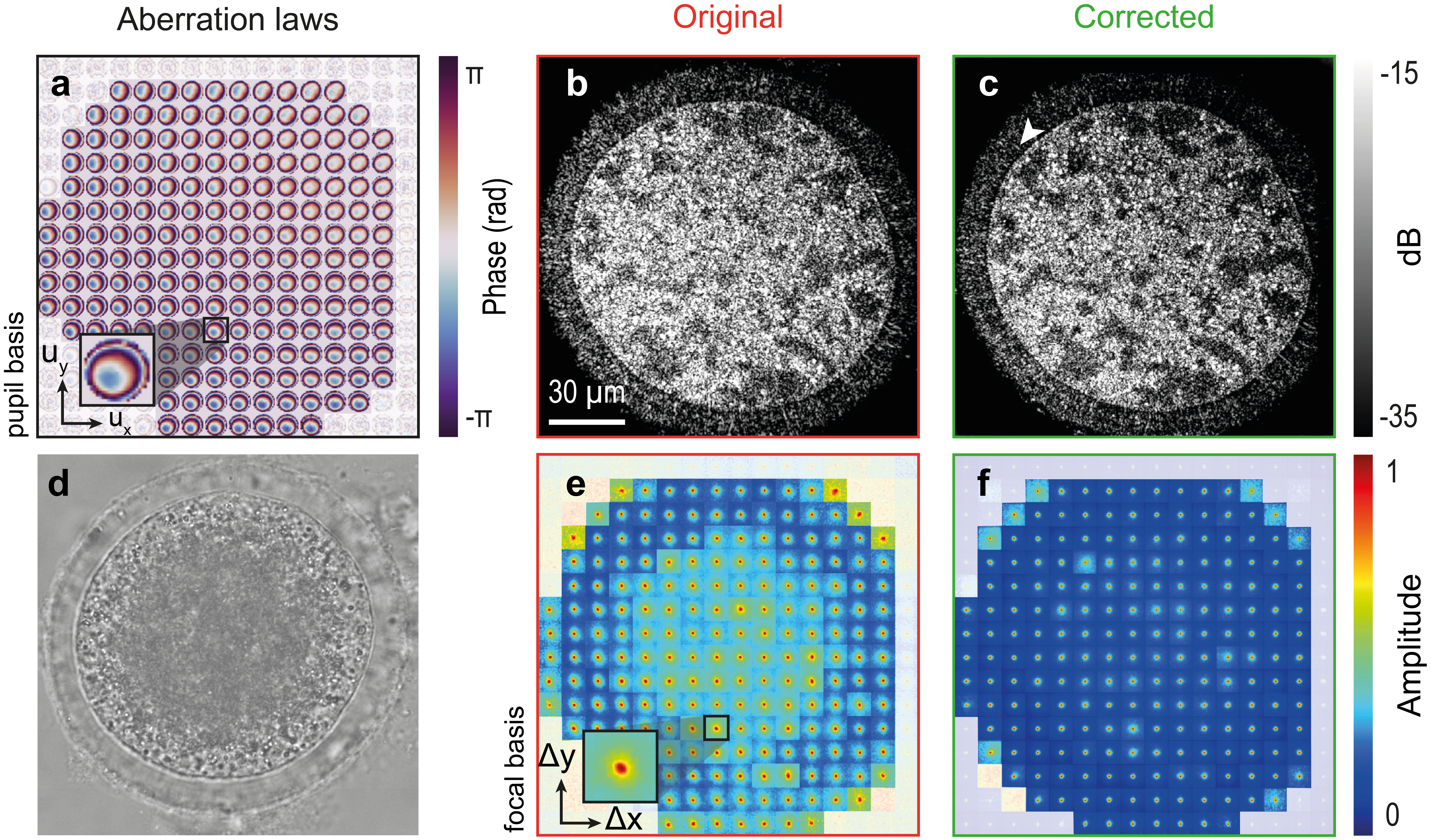}
\caption{\label{ext_fig1}\textbf{Optical matrix imaging of a denuded oocyte at depth {$z=70$ $\mu$m}}. \textbf{a}, Spatial map of the estimated aberration phase laws across the field-of-view. \textbf{b}-\textbf{c}, Transverse cross-sections of the oocyte before (\textbf{b}) and after (\textbf{c}) the computational aberration correction process (scale bar: 30 $\mu$m). Each image is normalized by its respective maximum intensity. The white arrow in c highlights the successful resolution of trans-zonal projections (TZPs), which are obscured by aberrations in the raw confocal image. \textbf{d}, Bright-field image of the oocyte. \textbf{e}-\textbf{f}, Maps of the reflection point-spread function (RPSF) before (\textbf{e}) and after (\textbf{f}) the RMI-based correction, illustrating the restoration of near-diffraction-limited resolution and the suppression of the multiple scattering background.}
\end{figure*}

\begin{figure}[ht]
\centering
\includegraphics[width=\textwidth]{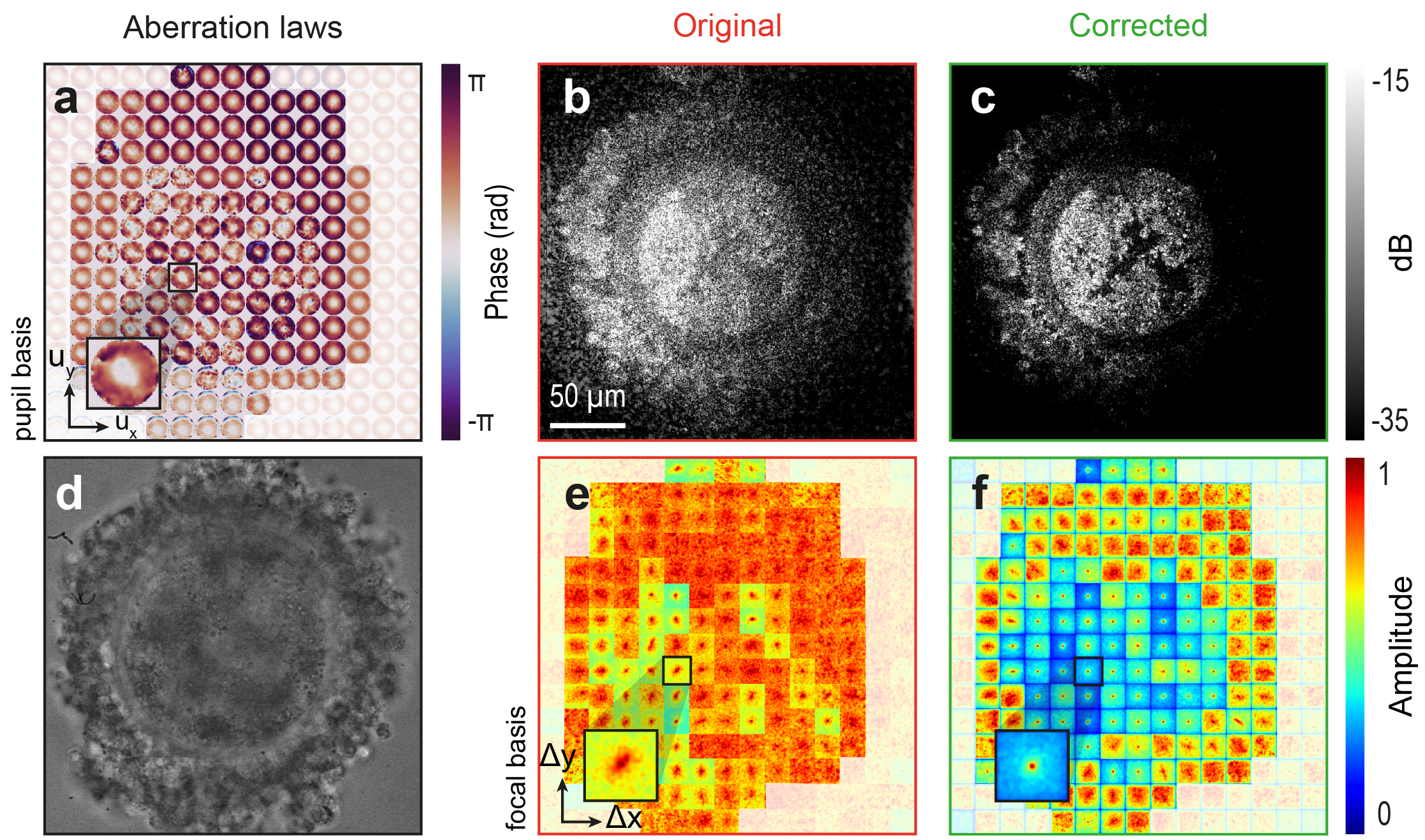}
\caption{\textbf{Optical matrix imaging of a cumulus-oocyte complex at depth $z=120 $ $\mu$m}. \textbf{a}, Spatial map of the estimated aberration phase laws across the field-of-view. \textbf{b}-\textbf{c}, En-face images of the oocyte before (\textbf{b}) and after (\textbf{c}) the computational aberration correction process (scale bar: 50 $\mu$m). Each image is normalized by its respective maximum intensity. \textbf{d}, Bright-field image of the oocyte.  \textbf{e}-\textbf{f}, Maps of the reflection point-spread function (RPSF) before (\textbf{e}) and after (\textbf{f}) the RMI-based correction, illustrating the restoration of near-diffraction-limited resolution and the suppression of the multiple scattering background.}
\label{ext_fig2}
\end{figure}

\begin{figure}[ht]
\centering
\includegraphics[width=\textwidth]{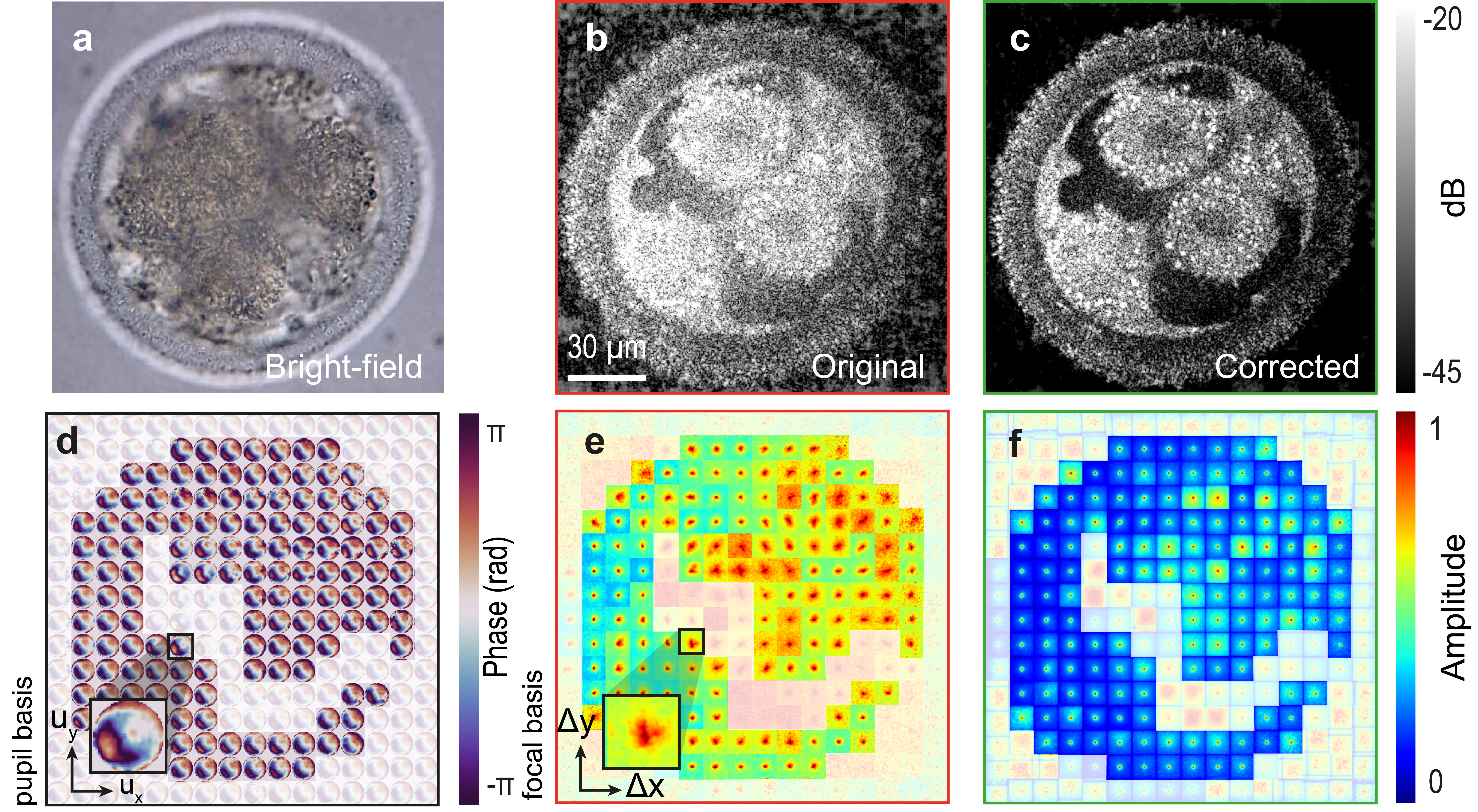}
\caption{\textbf{Optical matrix imaging of a {day 2} bovine embryo.} \textbf{a}, Bright field image of the embryo (scale bar: 30 $\mu$m). \textbf{b}-\textbf{c}, En-face images of the embryo at depth $z=98$ $\mu$m before (\textbf{b}) and after (\textbf{c}) after the computational aberration correction process (scale bar: 30 $\mu$m). Each image is normalized by its respective maximum intensity. \textbf{d}, {Map of aberration laws in a longitudinal $(x,z)$-plane.} \textbf{e}-\textbf{f} RPSF map in the same plane, before  (\textbf{e}) and after (\textbf{f}) aberration correction, respectively.}
\label{fig5ext}
\end{figure}

\begin{figure}[ht]
\centering
\includegraphics[width=\textwidth]{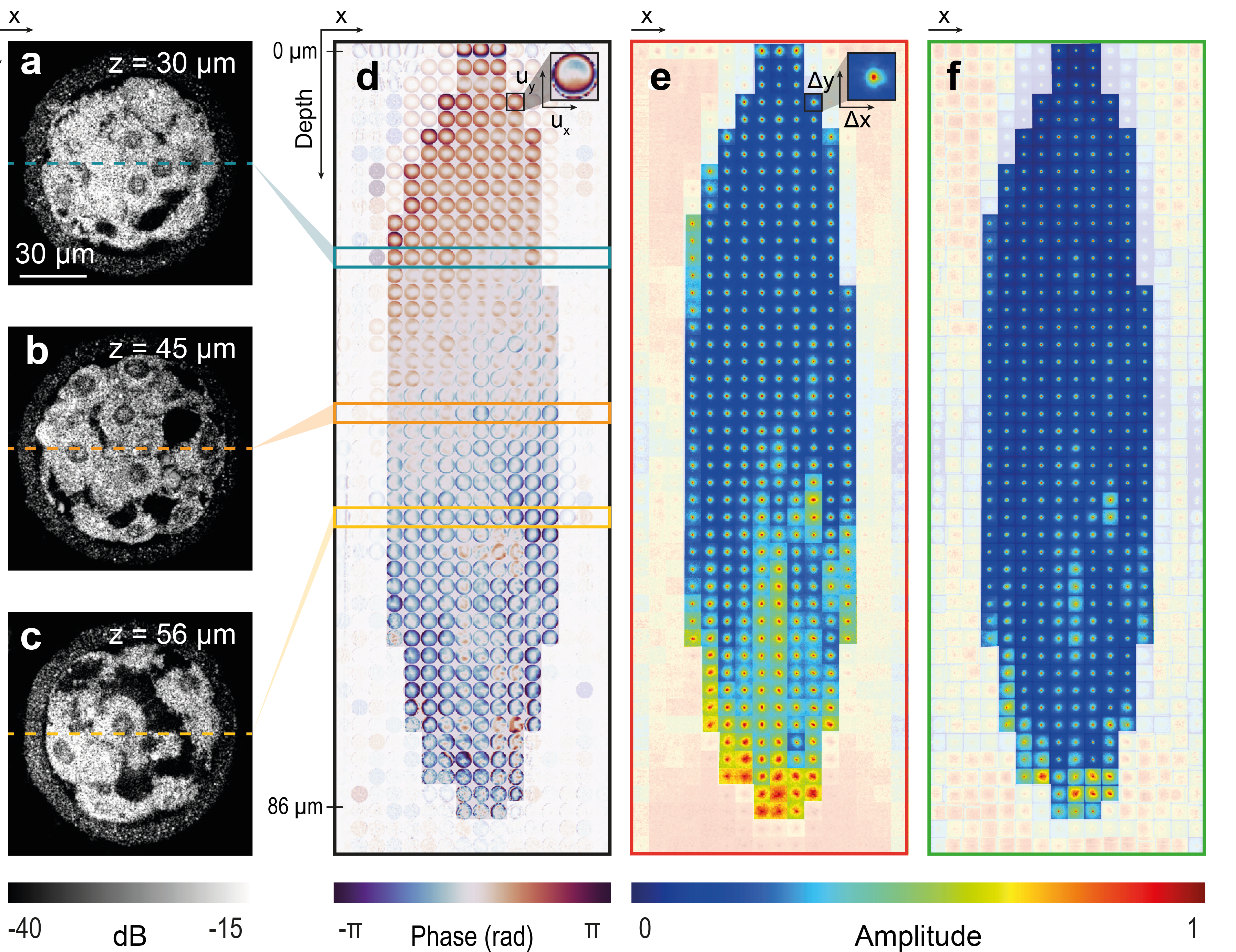}
\caption{\textbf{Reflection matrix imaging of a mouse morula.} \textbf{a}-\textbf{c}, Transverse cross-sections after aberration compensation at three different depths (scale bar: 30 $\mu$m). Each cross-section is normalized by its respective maximum intensity. \textbf{d}, {Map of aberration laws in a longitudinal $(x,z)$-plane.} \textbf{e}-\textbf{f}, RPSF map in the same plane, before (\textbf{e}) and after (\textbf{f})  aberration correction, respectively.}
\label{fig7}
\end{figure}

\end{document}